\documentclass{article}

\usepackage[preprint]{neurips_2025}

\usepackage[utf8]{inputenc} 
\usepackage[T1]{fontenc}    

\usepackage{url}            
\usepackage{booktabs}       
\usepackage{amsfonts}       
\usepackage{nicefrac}       
\usepackage{microtype}      
\usepackage{amsmath}        
\usepackage{graphicx}       
\usepackage{multirow}       
\usepackage{wrapfig}        
\usepackage{amssymb}        
\usepackage{subcaption}     
\usepackage{afterpage}      
\usepackage{siunitx}        
\usepackage{array}          
\usepackage{enumitem}       
\usepackage{algorithmic}    

\usepackage[dvipsnames,table]{xcolor} 
\definecolor{MyGreen}{RGB}{0, 165, 81}
\definecolor{MyNavyBlue}{RGB}{0, 111, 185}
\definecolor{MyRed}{RGB}{236, 31, 77}

\makeatletter
\providecommand{\adl@hline}{\hline}         
\providecommand{\adl@connect}{\cline{1-1}}  
\makeatother

\usepackage[most]{tcolorbox}    
\usepackage{adjustbox}          
\usepackage{marvosym}           

\usepackage{hyperref}
\hypersetup{
    colorlinks=true,
    urlcolor=black,
    citecolor=MyGreen,
    linkcolor=MyRed  
}

\newcommand{\inlineicon}[2][height=0.9em]{\raisebox{-0.2ex}{\includegraphics[#1]{#2}}}

\title{Verifiably Forgotten? Gradient Differences Still Enable Data Reconstruction in Federated Unlearning}

\vspace{-2 mm}

\author{
  Fuyao Zhang\textsuperscript{1, 2}\quad 
  Wenjie Li\textsuperscript{1, 2, \Letter}\quad 
  Yurong Hao\textsuperscript{1, 3}\quad 
  Xinyu Yan\textsuperscript{1} \quad\\
  \textbf{Yang Cao}\textsuperscript{\textbf{4}} \quad 
  \textbf{Wei Yang Bryan Lim}\textsuperscript{\textbf{1}}\\
    \textsuperscript{1} Nanyang Technological University\quad
    \textsuperscript{2} Xidian University\\
    \textsuperscript{3} Beijing Jiaotong University\quad
    \textsuperscript{4} Institute of Science Tokyo\\
    \texttt{\{hi.fyzhang, tom643190686\}@gmail.com}\quad
    \texttt{yurong.hao@bjtu.edu.cn}\\
    \texttt{xinyu020@e.ntu.edu.sg}\quad    \texttt{cao@c.titech.ac.jp}\quad
    \texttt{bryan.limwy@ntu.edu.sg}\\
}

\begin{document}

\maketitle
\vspace{-3 mm}
\begin{abstract}
Federated Unlearning (FU) has emerged as a critical compliance mechanism for data privacy regulations, requiring unlearned clients to provide verifiable Proof of Federated Unlearning (PoFU) to auditors upon data removal requests. However, we uncover a significant privacy vulnerability: when gradient differences are used as PoFU, \textit{honest-but-curious} auditors may exploit mathematical correlations between gradient differences and forgotten samples to reconstruct the latter. Such reconstruction, if feasible, would face three key challenges: (i) restricted auditor access to client-side data, (ii) limited samples derivable from individual PoFU, and (iii) high-dimensional redundancy in gradient differences. To overcome these challenges, we propose \textbf{I}nverting \textbf{G}radient difference to \textbf{F}orgotten data (IGF), a novel learning-based reconstruction attack framework that employs Singular Value Decomposition (SVD) for dimensionality reduction and feature extraction. IGF incorporates a tailored pixel-level inversion model optimized via a composite loss that captures both structural and semantic cues. This enables efficient and high-fidelity reconstruction of large-scale samples, surpassing existing methods. To counter this novel attack, we design an orthogonal obfuscation defense that preserves PoFU verification utility while preventing sensitive forgotten data reconstruction. Experiments across multiple datasets validate the effectiveness of the attack and the robustness of the defense. The code is available at \href{https://anonymous.4open.science/r/IGF}{\textcolor{MyNavyBlue}{\texttt{https://anonymous.4open.science/r/IGF}}}.
\end{abstract}

\section{Introduction}

\begin{wrapfigure}{r}{5.5 cm}
\vspace{-1.5 em}
\includegraphics[width=5.5 cm,height=4.4 cm]{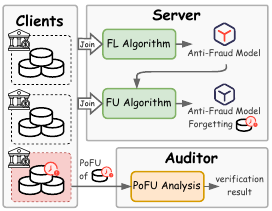}
\caption{Audition in verifiable FU}
\vspace{-1 em}
\label{fig:audit}
\end{wrapfigure}

The widespread adoption of Federated Learning (FL) enables distributed entities, such as financial institutions, healthcare providers, and IoT networks, to collaboratively train models without sharing raw data. This decentralized approach mitigates risks associated with data transfer, enhancing privacy and security for data owners. 
However, compliance with regulations like the \textit{right to be forgotten} under the General Data Protection Regulation (GDPR)~\cite{rosen2011right,pardau2018california} requires FL systems to remove specific data contributions from the global model and demonstrate that the model no longer depends on those data samples. 
Simply preventing raw data leaks is no longer sufficient to meet compliance requirements. This challenge has spurred the development of verifiable Federated Unlearning (FU)~\cite{liu2020federated}, a paradigm designed to verifiably forget the contribution of designated data from trained models.


Figure~\ref{fig:audit} illustrates a typical scenario where multinational financial institutions, acting as FL clients \inlineicon{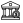}, collaboratively train an anti-fraud model~\cite{lindstrom2024federated}. Subsequently, the auditor mandates all clients to forget the outdated transaction data~\inlineicon{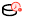} using the FU algorithm and receives the proof of FU (PoFU)~\cite{gao2024verifi,10416191,10937129,salem2020updates} from the unlearned client. Since the auditor cannot directly access the raw client data, they rely on PoFU, typically based on gradient differences between the original and unlearned models, as a non-invasive auditing tool. 
However, most research~\cite{wu2022federated,wang2022federated,zhong2025unlearning} primarily focuses on FU algorithm design, overlooking vulnerabilities to reconstruction attacks by third-party auditors~\cite{boenisch2023curious,le2023privacy}, particularly when gradient differences serve as PoFU. 

Recent advances in reconstruction attacks have exposed critical vulnerabilities in centralized machine (un)learning. For instance, DLG~\cite{zhu2019deep}, demonstrated that shared gradients can be exploited to reconstruct original training data, while subsequent work~\cite{geiping2020inverting} highlighted privacy leakage risks from gradient sharing. More recently, unlearning inversion attacks~\cite{hu2024learn} reconstruct forgotten data by only accessing the parameter deviations of the original and unlearned models. However, these approaches face three primary limitations when applying to federated unlearning scenarios: (i) they require \textit{white-box access} to calculate parameter deviations, (ii) they struggle with \textit{large-scale} data reconstruction due to the limited information encoded in these deviations, and (iii) the \textit{high dimensionality} of model parameters or gradients increases the computational cost of inversion models. 
More crucially, FU introduces additional complexities for reconstruction attack, the auditor lacks access to client-side raw data~\cite{279996} and relies solely on PoFU to evaluate unlearning efficacy. 
Current reconstruction attacks target model parameters or gradients, posing significant threats to machine unlearning in adversarial settings. However, reconstruction attacks exploiting gradient differences as PoFU remain underexplored. This gap motivates our research question:

\textit{\textbf{Q: Can gradient differences, used as PoFU, enable third-party auditors to reconstruct forgotten data? If so, how can high-fidelity, large-scale data reconstruction be achieved against high-dimensional gradient differences?}}

To address this, we propose a learning-based reconstruction attack in verifiable FU, named \textbf{I}nverting \textbf{G}radient difference to \textbf{F}orgotten data (IGF). To handle high-dimensional gradient differences, we employ Singular Value Decomposition (SVD) for dimensionality reduction, extracting essential features while eliminating redundancy, thus reducing the input dimensionality of the inversion model. We then design a pixel-level convolutional inversion model that learns the latent mapping between gradient differences and original data samples, optimized via a composite loss function incorporating structural and perceptual factors. This model efficiently reconstructs batches of forgotten samples from individual PoFU, bypassing per-sample optimization inefficiencies. These components synergize to enable robust, large-scale sample reconstruction across diverse datasets and global model architectures. Our main contributions are: 
\begin{itemize}
    \item We identify gradient differences used as PoFU as a novel attack surface for reconstruction attacks. By formalizing an \textit{honest-but-curious} third-party auditor, we demonstrate that passive observers can reconstruct forgotten samples~\cite{li2022auditing}. 
    \item We develop IGF attack framework, integrating SVD with a composite loss-optimized and pixel-level inversion network, achieving high reconstruction fidelity and computational efficiency.  
    \item Our experiments demonstrate that IGF outperforms the state-of-the-art GIAMU~\cite{hu2024learn} method, reducing reconstruction MSE by over 88\% and improving LPIPS by approximately 33\% on CIFAR-10, highlighting its effectiveness in federated unlearning.
    \item We further propose an orthogonal obfuscation defense mechanism to mitigate IGF, and validate both attack and defense efficacy through extensive experiments on public benchmark datasets.
\end{itemize}

\section{Related Work}
\textbf{Federated Unlearning.}
Federated unlearning has recently emerged to address the challenge of selectively removing specific clients or data points from a trained FL model.
This problem is motivated by regulatory requirements (e.g., the \textit{right to be forgotten} under GDPR) and the dynamic nature of real-world FL systems. Existing approaches can be categorized into two main types: \textit{Exact Federated Unlearning (EFU)}~\cite{liu2022right} and \textit{Approximate Federated Unlearning (AFU)}~\cite{halimi2207federated}. EFU achieves complete removal by retraining the model from scratch using the remaining data, ensuring that the influence of the target data is entirely eliminated. However, this method is computationally intensive and may not be practical for large-scale FL systems.
AFU methods aim to reduce computational overhead by approximating the unlearning process through applying gradient ascent to maximize the loss. Among approximate methods, Wang et al.~\cite{wang2024fedu} proposes that clients estimate the gradient influence of the data to be removed using local remaining data and then apply gradient ascent to negate this influence. A subsequent fine-tuning step is introduced to preserve overall utility. Similarly, Xu et al.~\cite{xu2024update} employ model explanations to identify key parameter channels associated with the forgotten categories, and update only those channels in reverse. Meanwhile, Gu et al.~\cite{gu2024unlearning} pre-generates linear transformation parameters related to the target data during the training phase and applies reverse transformations to eliminate unwanted effects. The above methods balance effectiveness and efficiency. 
Some works~\cite{chen2025fedmua,wang2025poisoning} explore how to diminish the model's utility by poisoning or cause excessive forgetting through malicious requests, but overlook potential reconstruction vulnerabilities during the verification stage.

\textbf{Gradient Inversion Attack.}
Recent studies have leveraged gradient inversion techniques to reconstruct clients' private training data in FL~\cite{zhang2023generative, jeon2021gradient, fang2023gifd, sun2024gi, wu2023learning}. Zhang et al.~\cite{zhang2023generative} demonstrate the feasibility of generative gradient inversion in FL by constructing an over-parameterized convolutional neural network that satisfies gradient-matching requirements. Similarly, Jeon et al.~\cite{jeon2021gradient} leverages pre-trained generative models as priors to circumvent direct optimization in high-dimensional pixel space and reconstructs data via latent-space parameter optimization. Additionally, Fang et al.~\cite{fang2023gifd} adopts a staged optimization strategy for the intermediate feature domains of generative models, progressively optimizing from the latent space to intermediate layers to enhance attack effectiveness. Sun et al.~\cite{sun2024gi} introduces an anomaly detection model to capture latent distributions from limited data, using it as a regularization term to enhance attack performance. In the context of FU, Hu et al.~\cite{hu2024learn} reveal the feature and label information by analyzing differences between the original and unlearned models. 

Therefore, traditional gradient inversion attacks focus on reconstructing training data directly from original gradients provided by clients in standard federated learning scenarios. In contrast, our work targets \textbf{gradient differences} used as PoFU, where the attacker must reconstruct deleted data from indirect and variant gradient information. This introduces unique challenges, gradient differences contain limited and mixed signals with weaker correlations to the forgotten samples, requiring fundamentally different inversion approaches. 

\section{Methodology}

\subsection{Problem Formulation}

\textbf{Federated Learning (FL).} 
In the FL framework with $H$ clients, each client $i$ ($i \in [H]$) holds a local dataset $\mathcal{D}_i$ containing $|\mathcal{D}_i|$ samples. Let $\mathbf{M}$ denote the original global model parameterized by $\boldsymbol{\theta}$, and consider a supervised learning objective that minimizes the empirical loss over the federated dataset $\mathcal{D} = \bigcup_{i=1}^H \mathcal{D}_i$:$\mathcal{L}(\boldsymbol{\theta}) = \frac{1}{|\mathcal{D}|} \sum_{(x,y) \in \mathcal{D}} \ell\big(\mathbf{M}(x; \boldsymbol{\theta}), y\big) $. The stochastic gradient for a data sample $(x_s, y_s) \in \mathcal{D}$ is $\mathbf{g}_s = \nabla_{\boldsymbol{\theta}} \ell\big(\mathbf{M}(x_s; \boldsymbol{\theta}), y_s\big)$. Federated Averaging (FedAvg)~\cite{mcmahan2017communication} operates through $T$ global rounds. At global round $t \in [T]$, the server broadcasts the current global model parameters $\boldsymbol{\theta}^t$ to all clients. Each client $i$ updates $\boldsymbol{\theta}^t$ via local SGD on $\mathcal{D}_i$: $\boldsymbol{\theta}_i^t = \boldsymbol{\theta}^t - \eta \cdot \nabla_{\boldsymbol{\theta}} \mathcal{L}_i(\boldsymbol{\theta}^t)$, where $\mathcal{L}_i(\boldsymbol{\theta}^t) = \frac{1}{|\mathcal{D}_i|} \sum_{(x,y) \in \mathcal{D}_i} \ell\big(\mathbf{M}(x; \boldsymbol{\theta}^t), y\big)$. Server aggregates via weighted averaging:
\begin{equation}
      \boldsymbol{\theta}^{t+1} = \sum_{i=1}^H \frac{|\mathcal{D}_i|}{|\mathcal{D}|} \boldsymbol{\theta}_i^t, \quad |\mathcal{D}| = \sum_{i=1}^H |\mathcal{D}_i|.
\end{equation}
The final global model after $T$ rounds is $\boldsymbol{\theta}^T$.

\textbf{FU Scenarios.}  
Let $\mathcal{C}_n \subseteq [H]$ denote clients retaining their original datasets $\{\mathcal{D}_j\}_{j \in \mathcal{C}_n}$, and $\mathcal{C}_u$ represent unlearned clients modifying their local datasets $\{\mathcal{D}_i\}_{i \in \mathcal{C}_u}$. Following~\cite{zhong2025unlearning}, we formalize three scenarios:
(i) \textit{sample-level unlearning:} For each client $i \in \mathcal{C}_u$, partition $\mathcal{D}_i$ into retained $\mathcal{D}_{i}^r$ and forgotten subsets $\mathcal{D}_{i}^f = \mathcal{D}_{i}\setminus \mathcal{D}_{i}^r$; 
(ii) \textit{class-level unlearning:} Each client $i \in \mathcal{C}_u$ removes all samples of target class $y^f$, yielding $\mathcal{D}_i^f = \{(x,y) \in \mathcal{D}_i \mid y = y^f\}$ with $\mathcal{D}_i^r = \mathcal{D}_i \setminus \mathcal{D}_i^f$; 
(iii) \textit{client-level unlearning:} Each client $i \in \mathcal{C}_u$ sets $\mathcal{D}_i^f = \mathcal{D}_i$ and $\mathcal{D}_i^r = \emptyset$. We denote the unlearned global model as $^u\mathbf{M}$, the forgotten dataset as $\mathcal{D}^{\text{forgotten}} = \bigcup_{i \in \mathcal{C}_u} \mathcal{D}_i^f $ and the retained dataset as $\mathcal{D}^{\text{retained}} = ( \bigcup_{j \in \mathcal{C}_n} \mathcal{D}_j ) \cup ( \bigcup_{i \in \mathcal{C}_u} \mathcal{D}_i^r )$.

\textbf{FU Methods.} We implement two mainstream FU approaches: 
(i) \textbf{EFU} retrains the global model on dataset $\mathcal{D}^{\text{retained}}$ from scratch, minimizing $\sum_{(x,y) \in \mathcal{D}^{\text{retained}}} \ell(\mathbf{M}(x; \boldsymbol{\theta}), y)$. This method precisely removes contributions of $\mathcal{D}^{\text{forgotten}}$ from the global model.
(ii) \textbf{AFU} performs projected gradient ascent and constrains maximization on $\mathcal{D}^{\text{forgotten}}$. For each client $i \in \mathcal{C}_u$, it computes $\boldsymbol{\theta}^{\prime}_i = \boldsymbol{\theta}^{T} + \eta_u \cdot \nabla_{\boldsymbol{\theta}} \mathcal{L}^{\prime}_i(\boldsymbol{\theta}^{T})$
where
$\mathcal{L}^{\prime}_i(\boldsymbol{\theta}^{T}) = \frac{1}{|\mathcal{D}^f_i|} \sum_{(x,y) \in \mathcal{D}^f_i} \ell\big(^u\mathbf{M}(x; \boldsymbol{\theta}^{T}), y\big)$
but maintains $\|\boldsymbol{\theta}^{\prime}_i - \boldsymbol{\theta}^{T}\|_2 \leq \zeta$, where $\zeta$ is the parameter deviation constraint. Then the server aggregates the unlearned local model parameters:
\begin{equation}
      \boldsymbol{\theta}^{\prime} = \sum_{i \in \mathcal{C}_u} \frac{|\mathcal{D}^f_i|}{|\mathcal{D}^{\text{forgotten}}|} \boldsymbol{\theta}_i^{\prime}, \quad 
      |\mathcal{D}^{\text{forgotten}}| = \sum_{i \in \mathcal{C}_u} |\mathcal{D}^f_i|,
\end{equation} 
and fine-tunes $^u\mathbf{M}$ with $\boldsymbol{\theta}^{\prime}$ on $\mathcal{D}^{\text{retained}}$.

\textbf{Verification in FU.} Each unlearned client $i \in \mathcal{C}_u$ locally computes PoFU of gradient differences $\mathbf{\Delta g}^{(n_i)} = \left\{ \mathbf{\Delta g}_j^{(n_i)} = \nabla_{\boldsymbol{\theta}} \ell\big(\mathbf{M}(x_j; \boldsymbol{\theta}^T), y_j\big) - \nabla_{\boldsymbol{\theta}} \ell\big(^u\mathbf{M}(x_j; \boldsymbol{\theta^{\prime}}), y_j\big) | {(x_j,y_j) \in \mathcal{D}_i^f} \right\}$. 
Auditor receives PoFUs $\mathbf{\Delta g}  = \{ \mathbf{\Delta g}^{(n_i)} \}_{i \in \mathcal{C}_u}$ and validates unlearning by checking each $\|\mathbf{\Delta g}_j^{(n_i)}\|_2 \leq \tau$ with predefined threshold $\tau$. 
The necessity of the gradient differences in verifiable FU lies in ensuring that a data point $(x,y)$ is included in the training dataset of the original model $\mathbf{M}$ but excluded from that of the unlearned model $^u\mathbf{M}$.

\textbf{Threat Assumption.} 
We model the auditor, denoted $\mathcal{A}$, as an \textit{honest-but-curious} entity that strictly follows the FU protocol but seeks to infer private client data. Consistent with prior reconstruction attacks~\cite{wu2023learning,hu2024learn,geiping2020inverting,zhu2019deep}, $\mathcal{A}$ possesses an auxiliary dataset $\mathcal{D}^{\text{aux}}$. Operating in a gray-box setting, $\mathcal{A}$ lacks knowledge of the global model's architecture but can collude with the server to query the flattened gradient for arbitrary samples from both the original model $\mathbf{M}$, and the unlearned model $^u\mathbf{M}$. During the exploitation phase, $\mathcal{A}$ passively collects PoFUs $\mathbf{\Delta g}$, and endeavors to reconstruct the forgotten samples. 

\subsection{Framework of IGF}
\begin{figure}[]
    \centering
    \includegraphics[width=\linewidth]{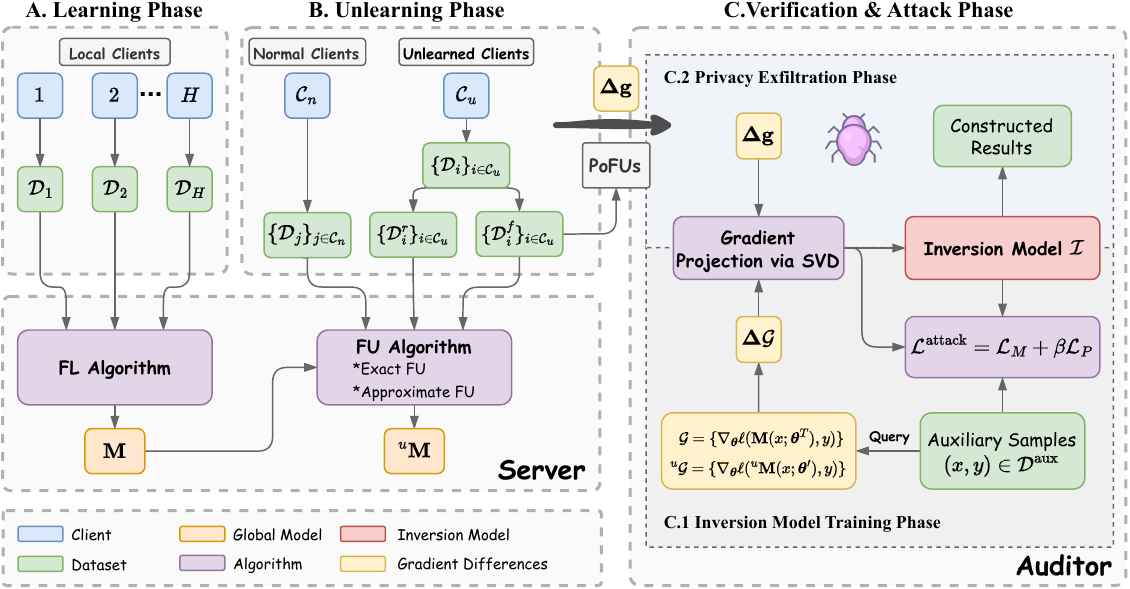}
    \caption{Schematic overview of IGF framework. 
    \textbf{A. Learning Phase}: Clients collaboratively train the global model via FL.
    \textbf{B. Unlearning Phase}: The unlearned clients are required to forget specific data contributions and submit the proof of federated unlearning (PoFU).
    \textbf{C. Verification \& Attack Phase}: The \emph{honest-but-curious} auditor verifies PoFUs, while attempting to infer forgotten data using a pre-trained inversion model $\mathcal{I}$.} 
    \label{main}
    \vspace{-2 em}
\end{figure}


We adopt a learning-based inversion model to invert gradient differences to forgotten samples during the verification phase of FU. The main schematic of IGF is shown in Figure~\ref{main}, and the formalized details are as follows:

\vspace{2mm}
\textbf{Inversion Model Training Phase.} (i) \textbf{Preparation of Training Dataset}. To prepare the training data for inversion model $\mathcal{I}$, for each data point $(x,y)$ in auxiliary dataset $\mathcal{D}^{\text{aux}}$, the auditor $\mathcal{A}$ collects gradients: 
\begin{align}
\left\{
\begin{aligned}
\mathcal{G} &= \{ \nabla_{\boldsymbol{\theta}}\ell(\mathbf{M}(x; \boldsymbol{\theta}^T), y) \}_{(x,y) \in \mathcal{D}^{\text{aux}}}\\
^u\mathcal{G} &= \{ \nabla_{\boldsymbol{\theta}}\ell(^u\mathbf{M} (x; \boldsymbol{\theta}^\prime), y)\}_{(x,y) \in \mathcal{D}^{\text{aux}}},
\end{aligned}
\right.
\end{align}
where $\mathcal{G}$ and $^u\mathcal{G}$ denote the sets of flatten gradients queried from $\mathbf{M}$ and $^u\mathbf{M}$, respectively. Gradient differences $\mathbf{\Delta} \mathcal{G} = \{ \mathcal{G}_i - ^u\mathcal{G}_{i} | (x_i,y_i) \in \mathcal{D}^{\text{aux}}\}$ form a set of $d$-dimensional vectors, with $d$ as the number of trainable parameters. 

\vspace{2mm}
(ii) \textbf{Gradient Differences Projection via SVD}. To extract the key features and address redundancy caused by the high dimensionality of gradient differences, $\mathcal{A}$ projects $\mathbf{\Delta} \mathcal{G}$ to a lower-dimensional space using SVD. Let the $m$ denote the number of samples in $\mathcal{D}^{\text{aux}}$, $\mathcal{A}$ constructs a matrix $\boldsymbol{\Psi} = [\mathbf{\Delta} \mathcal{G}_1^\top, \mathbf{\Delta} \mathcal{G}_2^\top, \ldots, \mathbf{\Delta} \mathcal{G}_m^\top] \in \mathbb{R}^{m \times d}$ from the gradient differences $\{\mathbf{\Delta} \mathcal{G}_i\}_{i=1}^m$ of the auxiliary dataset $\mathcal{D}^{\text{aux}}$, where $m \ll d$ typically holds. $\mathcal{A}$ centers the gradient differences by subtracting the mean vector $\boldsymbol{\mu} = \frac{1}{m}\sum_{i=1}^m \mathbf{\Delta} \mathcal{G}_i$, resulting in $\mathbf{\Delta} \mathcal{G}_i^{\text{cen}} = \mathbf{\Delta} \mathcal{G}_i - \boldsymbol{\mu}$ and the centered matrix $\mathbf{\Psi}^{\text{cen}}$. decomposes the centered matrix $\mathbf{\Psi}^{\text{cen}}$:
\begin{equation}
    \mathbf{\Psi}^{\text{cen}} = \mathbf{U} \mathbf{\Sigma} \mathbf{V}^\top,
\end{equation}

with $\mathbf{U} \in \mathbb{R}^{m \times m}$, $\mathbf{V} \in \mathbb{R}^{d \times d}$, and diagonal matrix $\mathbf{\Sigma}$ contains singular values $\sigma_1 \geq \sigma_2 \geq \dots \geq \sigma_m \geq 0$. To preserve essential information while reducing dimensionality, $\mathcal{A}$ selects the smallest $k$ such that the cumulative explained variance exceeds a threshold $\nu$:
\begin{equation}
k = \min\left\{ j \, \bigg| \, {\sum_{i=1}^j \sigma_i^2}/{\sum_{i=1}^m \sigma_i^2} \geq \nu \right\}.
\end{equation}
So $\mathcal{A}$ gets the projection matrix $\mathbf{V}^{[k]}$ denotes the first $k$ columns of $\mathbf{V}$. And the projected gradient differences of $\mathcal{D}^{\text{aux}}$ are computed as $\mathbf{\Delta} \mathcal{G}^{\text{proj}} = \boldsymbol{\Psi} \mathbf{V}^{[k]} \in \mathbb{R}^{m \times k}$.

\vspace{2mm}
(iii) \textbf{Training Inversion Model}.
$\mathcal{A}$ trains the inversion model, denoted as $\mathcal{I}$ and parameterized by $\boldsymbol{\omega}$, to map projected gradient differences to samples in $\mathcal{D}^{\text{aux}}$ by minimizing the composite loss function: 
\begin{equation}
\mathcal{L}^{\text{attack}}(\boldsymbol{\omega}) = \mathcal{L}_{M}(\boldsymbol{\omega}) + \beta \mathcal{L}_{P}(\boldsymbol{\omega}), 
\end{equation}
where $\beta$ trades off between pixel-level accuracy and perceptual quality. This design is common in image reconstruction tasks and can flexibly adjust the optimization objectives of the model to ensure that the reconstruction results are both accurate and natural. specifically, $\mathcal{L}_{M}$ quantifies the structural pixel-level discrepancy between reconstructed image $\mathcal{I}(\mathbf{\Delta} \mathcal{G}_i^{\text{proj}};\boldsymbol{\omega})$ and ground truth image $x_{i}$:
\begin{equation}
\mathcal{L}_{M}(\boldsymbol{\omega}) =  \frac{1}{m} \sum_{i=1}^{m} \| \mathcal{I}(\mathbf{\Delta} \mathcal{G}_i^{\text{proj}};\boldsymbol{\omega}) - {x}_{i}\|_{2}^{2}.
\end{equation}
Similarly, we define $\mathcal{L}_{P}$, which measures the semantic similarity between the reconstructed and true images using a VGG-based feature extractor $\phi(\cdot)$:
\begin{equation}
\mathcal{L}_{P}(\boldsymbol{\omega}) = 
\frac{1}{m} \sum_{i=1}^{m} \| \phi \left( \mathcal{I}(\mathbf{\Delta} \mathcal{G}_i^{\text{proj}};\boldsymbol{\omega}) \right) - \phi \left( {x}_{i}\right)\|_{2}^{2}
\end{equation}
Further, we elaborately designed the architecture of $\mathcal{I}$ to capture the latent mapping between gradient differences and images effectively. $\mathcal{I}$ employs a pixel-level convolutional network for progressive upsampling, which reduces artifacts in the reconstructed images. This design facilitates a nonlinear transformation from PoFU space to structured image space. Further architectural details are provided in Appendix~\ref{Inversion model architecture}. 

\textbf{Privacy Exfiltration Phase.}
Following the training phase, the auditor $\mathcal{A}$ possesses the projection matrix $\mathbf{V}^{[k]}$ and the inversion model $\mathcal{I}$ with parameter $\boldsymbol{\omega}$. Upon receiving PoFUs, 
for each PoFU $\mathbf{\Delta} \mathbf{g}^{(n_i)}$ of each client $i \in \mathcal{C}_u$, 
$\mathcal{A}$ constructs the matrix $\boldsymbol{\Psi}^{(n_i)} = \left[ \mathbf{\Delta} {\mathbf{g}^{(n_i)}_1}^\top, \mathbf{\Delta} {\mathbf{g}^{(n_i)}_2}^\top, \ldots, {\mathbf{g}^{(n_i)}_{n_i}}^\top \right] \in \mathbb{R}^{n_i \times d}$, 
where $n_i$ denotes the number of samples in $\mathcal{D}_i^f$. This matrix is then projected into a lower-dimensional space ${\mathbf{\Delta} \mathbf{g}^{(n_i)}}^{\text{proj}} = \boldsymbol{\Psi}^{(n_i)} \mathbf{V}^{[k]} \in \mathbb{R}^{n_i \times k}$. The batched reconstruction of projected gradient differences ${\mathbf{\Delta} \mathbf{g}^{(n_i)}}^{\text{proj}}$ is performed as follows:

\begin{equation}
\hat{\textbf{x}}^{(n_i)} = \{\hat{x}_{j} = \mathcal{I}({\mathbf{\Delta} \mathbf{g}_j^{(n_i)}}^{\text{proj}}; \boldsymbol{\omega}) | j \in [n_i]\},
\end{equation}

where $\hat{\textbf{x}}^{(n_i)}= \{\hat{x}_1, \hat{x}_2, \ldots, \hat{x}_{n_i}\}$ represents the $n_i$ reconstructed samples of client $i$. This exploitation enables $\mathcal{A}$ to utilize the pre-trained inversion model to implement the large-scale reconstructions from individual PoFU, thereby compromising data privacy even from the passive view.

\subsection{Orthogonal Obfuscation Defense Method}

\begin{wrapfigure}{r}{4.0cm}
\vspace{-2.2 em}
\includegraphics[width=4.0cm,height=4.8cm]{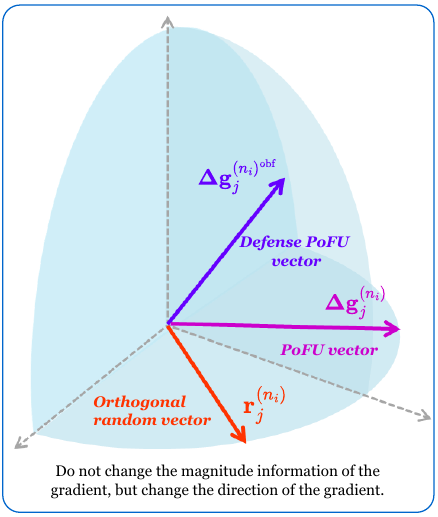}
\caption{Schematic of orthogonal obfuscation defense}
\vspace{-4 em}
\label{defense_main}
\end{wrapfigure}

Our inversion model exploits the directional information in gradient differences to reconstruct sensitive training data. Traditional defense methods often fail to disrupt the directional patterns, preserving the overall gradient differences structure and remaining susceptible to statistical recovery techniques. 
As illustrated in Figure~\ref{defense_main}, we propose a defense strategy that alters the vector direction while retaining the L2-norm information necessary for auditing. Our approach projects gradient differences into an orthogonal subspace, thereby disrupting the patterns and spatial structures that attackers rely on to reconstruct the forgotten sample. 

For each PoFU $\mathbf{\Delta} \mathbf{g}^{(n_i)}$ of unlearned client $i$, $i$ needs to modify the direction of each entry $\mathbf{\Delta} \mathbf{g}_j^{(n_i)}$ but maintain its L2-norm. We introduce random vectors $\mathbf{r}^{(n_i)}$ that are orthogonal to $\mathbf{\Delta} \mathbf{g}^{(n_i)}$ element-wisely. The construction begins by sampling an initial random vector $\mathbf{r}_j^{(n_i)}$ with the same dimensionality as $\mathbf{\Delta} \mathbf{g}_j^{(n_i)}$, drawn from a standard normal distribution $\mathbf{r}_j^{(n_i)} \sim \mathcal{N}(0, 1)^d$. Then client $i$ applies the Gram-Schmidt orthogonalization~\cite{orthogonalization_eom} to compute: 
\begin{equation}
\mathbf{\Delta} \mathbf{g}_j^{(n_i)^{\text{obf}}} = \mathbf{r}_j^{(n_i)} - \frac{\mathbf{r}_j^{(n_i)^{\top}} \mathbf{\Delta} \mathbf{g}_j^{(n_i)}}{\| \mathbf{\Delta} \mathbf{g}_j^{(n_i)} \|^2} \mathbf{\Delta} \mathbf{g}_j^{(n_i)}.
\end{equation}
This step ensures that $\mathbf{\Delta} \mathbf{g}_j^{(n_i)^{\text{obf}}}$ lies in a subspace orthogonal to $\mathbf{\Delta} \mathbf{g}_j^{(n_i)}$, effectively decoupling its direction from the original PoFU vector while preserving the randomness needed for obfuscation.

\section{Experiment}
\subsection{Experiment Settings}
\textbf{Datasets and Metrics.} 
We assess IGF framework on widely adopted benchmark datasets: CIFAR-10, CIFAR-100~\cite{krizhevsky2009learning}, MNIST~\cite{lecun1998gradient}, and Fashion-MNIST~\cite{xiao2017fashion}. These datasets offer diverse challenges, featuring varying image resolutions ($32\times32$ for CIFAR, $28\times28$ for MNIST and Fashion-MNIST) and class numbers ($10$ to $100$), making them an ideal testbed for assessing generalization. To measure the efficacy of our IGF, we employ established metrics for reconstruction attacks: Mean Squared Error (MSE), Peak Signal-to-Noise Ratio (PSNR), and Learned Perceptual Image Patch Similarity (LPIPS)~\cite{geiping2020inverting,hu2024learn,sun2024gi,zhang2018unreasonable}. Further details are provided in Appendix~\ref{Appendix_Datasets} and~\ref{appendix_metrics}.

\textbf{Models.} To investigate the impact of global model complexity on our attack, we adopt two architectures: a convolutional neural network (\textit{ConvNet}) and a deeper residual network (\textit{ResNet20})~\cite{he2016deep}. These are tested on CIFAR-10 and CIFAR-100, enabling us to probe the attack’s robustness across architectural variations and to explore how the proposed inversion model scales with the network complexity of the global model. 

\textbf{Training Setup.} In cross-silo FU and FL training, we configure $40$ clients with $10\%$ client selection and conduct $20$ global rounds to derive the original and unlearned models. For unlearning, we designate $1000$ samples to be forgotten, and our ablation experiments demonstrate that both in-distribution and out-of-distribution auxiliary data can effectively achieve attacking. We consider an \textit{honest-but-curious} adversary $\mathcal{A}$ capable of storing or collecting a small auxiliary dataset, with a size comparable to a typical validation or test set, consistent with prior work~\cite{sun2024gi, wu2023learning}. During the attack phase, we train the inversion model with a batch size of $256$, a learning rate of $10^{-4}$, and a fixed random seed of $1234$ for reproducibility. All experiments are implemented in PyTorch and executed on NVIDIA A10 GPUs, with each training run requiring approximately 1 hour.

\subsection{Experimental Results}This section presents comprehensive experiments validating the effectiveness of IGF framework across diverse datasets, FU scenarios (\textit{sample-level}, \textit{class-level}, \textit{client-level}), FU methods (EFU/AFU), and global model architectures. We report numerical results and visual reconstructions to demonstrate IGF’s high-fidelity reconstruction capabilities. Additionally, we evaluate IGF’s resilience against five common defense mechanisms and highlight the efficacy of our proposed orthogonal obfuscation defense. Finally, we conduct extensive ablation studies to assess the contributions of IGF’s key components. 

\begin{table}[]
    \caption{Reconstruction performance (MSE, PSNR, and LPIPS) on CIFAR-10 and CIFAR-100 datasets with \textit{ConvNet} and \textit{ResNet20} as global models. Gradient differences are applied with no defense. Each cell reports results for EFU / AFU, with \textbf{bold} indicating the best performance.}
  \label{table:result1}
  \centering
  \begin{adjustbox}{max width=\textwidth}
\begin{tabular}{cccccccccccc}
    \toprule
    \multirow{2}{*}{\textbf{Backbone}} & 
    \multirow{2}{*}{\textbf{Method}} & 
    \multirow{2}{*}{\textbf{FU Scenario}} & 
    \multicolumn{3}{c}{\textbf{CIFAR-10}} & 
    \multicolumn{3}{c}{\textbf{CIFAR-100}} \\
    \cmidrule(lr){4-6} \cmidrule(lr){7-9}
    & & & MSE $\textcolor{MyRed}{\downarrow}$ & PSNR $\textcolor{MyGreen}{\uparrow}$ & LPIPS $\textcolor{MyRed}{\downarrow}$ & MSE $\textcolor{MyRed}{\downarrow}$ & PSNR $\textcolor{MyGreen}{\uparrow}$ & LPIPS $\textcolor{MyRed}{\downarrow}$ \\
    \midrule
    \multirow{4}{*}{\textit{ConvNet}}
      & Ours & \textit{sample-level} & 0.0211 / \textbf{0.0218} & 17.19 / \textbf{17.09} & \textbf{0.3261} / 0.3624 & \textbf{0.0364} / \textbf{0.0261} & \textbf{14.97} / \textbf{16.07} & 0.4383 / \textbf{0.4190} \\
      & Ours & \textit{class-level}  & 0.0259 / 0.0234 & 16.08 / 16.51 & 0.3531 / 0.3316 & 0.0397 / 0.0298 & 14.41 / 15.73 & 0.4451 / 0.4201 \\
      & Ours & \textit{client-level} & \textbf{0.0206} / 0.0223 & \textbf{17.32} / 16.78 & 0.3747 / 0.3558 & 0.0382 / 0.0265 & 14.65 / \textbf{16.07} & \textbf{0.4361} / 0.4223 \\
      & GIAMU & \textit{sample-level} & 0.2330 / 0.2460 & 13.22 / 12.78 & 0.3390 / \textbf{0.3190} & -- & -- & -- \\
    \midrule
    \multirow{3}{*}{\textit{ResNet20}}
      & Ours & \textit{sample-level} & 0.0445 / 0.0564 & 14.05 / 13.02 & \textbf{0.4607} / 0.4719 & \textbf{0.0391} / \textbf{0.0353} & \textbf{14.56} / \textbf{15.02} & 0.4267 / 0.4025 \\
      & Ours & \textit{class-level}  & 0.0535 / \textbf{0.0512} & 13.01 / \textbf{13.21} & 0.4608 / \textbf{0.4366} & 0.0474 / 0.0438 & 13.49 / 13.84 & \textbf{0.4060} / 0.4032 \\
      & Ours & \textit{client-level} & \textbf{0.0435} / 0.0533 & \textbf{14.12} / 13.08 & 0.4617 / 0.4983 & 0.0422 / 0.0362 & 14.27 / 14.73 & 0.4187 / \textbf{0.3627} \\
    \bottomrule
  \end{tabular}
  \end{adjustbox}
  \vspace{-1 em}
\end{table}

\textbf{Reconstruction Performance across Datasets.} The results presented in Table~\ref{table:result1} provide compelling evidence of IGF's capability to reconstruct forgotten data with high fidelity. On CIFAR-10 with \textit{ConvNet} under EFU at the \textit{sample-level} ($1000$ forgotten samples), IGF achieves an MSE of $0.0211$, PSNR of $17.1947$, and LPIPS of $0.3261$, reflecting reconstructions with minimal pixel-wise errors and superior perceptual quality. On more challenging CIFAR-100, which contains $100$ classes instead of $10$, we observe a slight performance degradation, with MSE rising to $0.0364$, PSNR dropping to $14.9658$, and LPIPS increasing to $0.4383$. This decline is expected, as greater dataset complexity and inter-class variability heighten the difficulty of inversion. Nevertheless, IGF still demonstrates significant robustness, which can be attributed to our SVD-based projection technique and a tailored inversion model. 

\textbf{Adaptability across Global Model Architectures.} IGF also demonstrates adaptability across model architectures. On CIFAR-10, reconstruction performances with \textit{ConvNet} are slightly better than \textit{ResNet20} , with MSE values of $0.0211$ and $0.0445$, respectively. 
This gap stems from the increased complexity and depth of \textit{ResNet20}, which leads to more complex gradient patterns that complicate inversion. 
Despite this, IGF still achieves satisfactory reconstruction quality even with the deeper \textit{ResNet20} architecture, demonstrating the adaptability to diverse model architectures. 

\textbf{Adaptability across FU Scenarios.} We test IGF under three FU scenarios: \textit{sample-level unlearning} (the number of samples to be forgotten is set to 1000), \textit{class-level unlearning} (the class index to be forgotten is set to $1$), and \textit{client-level unlearning} (all samples from the third client are set to be forgotten). IGF exhibits stable performance, with \textit{ConvNet}’s MSE fluctuating within $0.0053$ under EFU on CIFAR-10, indicating resilience to differing unlearning granularity. In other configurations, alterations to the FU scenarios have a negligible impact on reconstruction performance, further highlighting IGF's stability.

\textbf{Vulnerability Comparison of FU Methods.} 
Experimental results reveal certain gaps in vulnerability to reconstruction attacks between EFU and AFU methods.
EFU outperforms AFU in reconstruction metrics on CIFAR-10 with \textit{ResNet20}, as EFU’s retraining from scratch yields clearer gradient differences reflecting forgotten data’s impact. In contrast, AFU’s gradient ascent operation introduces noise, complicating reconstruction. Despite this, IGF achieves reasonable reconstruction quality under AFU, highlighting a critical privacy risk: even approximate unlearning methods remain vulnerable to reconstruction attacks. 

\textbf{Comparison with Basline~\cite{hu2024learn}.} As shown in Table~\ref{table:result1}, we also compare our IGF framework with GIAMU~\cite{hu2024learn}, a recent gradient inversion attack for centralized machine unlearning. All results associated with GIAMU are sourced directly from~\cite{hu2024learn}. For \textit{sample-level} unlearning on CIFAR-10 under EFU, IGF outperforms GIAMU by 88.1\%, 30.1\%, and 3.8\% in MSE, PSNR, and LPIPS, respectively. Under AFU, gains are even more pronounced, with MSE and LPIPS improvements of 91.1\% and 33.6\%, respectively. Notably, few baseline attacks are suitable for direct comparison in FU reconstruction, as most existing methods target gradients from fully trained models, which are ill-suited for unlearning scenarios where models are modified to forget specific data. Moreover, IGF scales to reconstruct hundreds of times more samples than GIAMU, despite FU’s distributed nature, which restricts data access and complicates attacks compared to GIAMU’s white-box setting. 

\begin{figure}[!htbp]
    \centering
    \includegraphics[width=\textwidth]{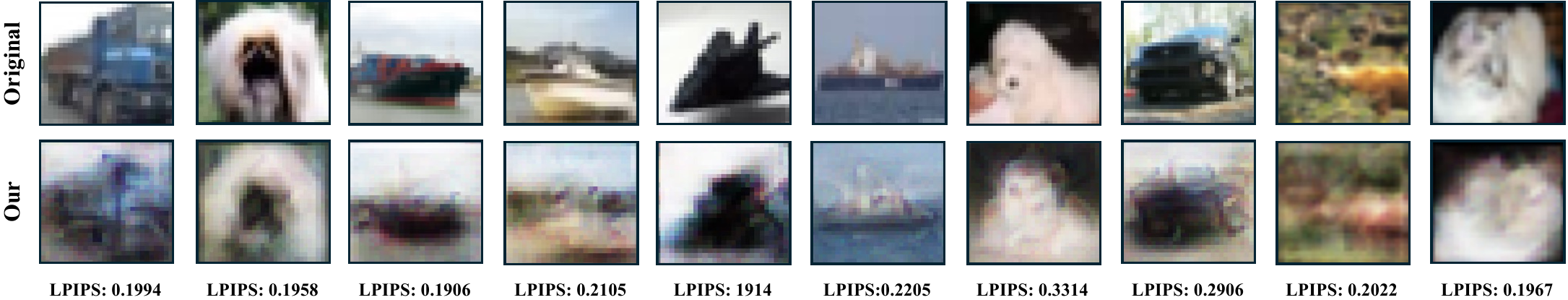}
    \caption{Original forgotten images and our reconstructed images on the CIFAR-10 dataset when the number of forgotten samples is 1000.}
    \label{reconstruct-img1}
    \vspace{-1mm}
\end{figure}

\textbf{Visual Inspection of Reconstructed Images.} Beyond quantitative metrics, visual inspection of the reconstructed images in Figure~\ref{reconstruct-img1} provides additional insights into the effectiveness of IGF. The reconstructed images clearly capture the essential features of the original forgotten samples, including object shapes, colors, and textures. This visual similarity reinforces the quantitative results and demonstrates that our attack can reconstruct forgotten data with sufficient fidelity to pose a genuine privacy risk. In addition, we extend IGF to MNIST and Fashion-MINST, which are composed of different image sizes from CIFAR. The reconstructed results, as shown in Figure~\ref{reconstruct-mnist}, show that reconstructed images are almost indistinguishable from the original images based on the gradient differences. The high-quality reconstruction is achieved through our composite optimization approach, which combines $\mathcal{L}_M$ with $\mathcal{L}_P$ loss. This combination ensures that the reconstructed images not only match the original images at the pixel level but also maintain perceptual similarity in terms of high-level features.

\begin{table}[!htbp]
  \vspace{-1 em}
  \caption{Reconstruction performance across three metrics on five common defense mechanisms.}
  \label{defensetable}
  \centering
  \begin{adjustbox}{max width=\textwidth}
  \begin{tabular}{cccccccc}
    \toprule
    \multirow{2}{*}{\textbf{Defense Method}} & \multicolumn{3}{c}{\textbf{Gradient Pruning}} & \textbf{Sign Compression} & \textbf{Gauss Noise} & \textbf{Gradient Perturb} & \textbf{Gradient Smooth} \\
    \cmidrule(r){2-4}
    & 0.7 & 0.8 & 0.9 & 0.001 & 0.1 & 0.01 & 0.1 \\
    \midrule
    MSE $\textcolor{MyRed}{\downarrow}$  & 0.0216 & 0.0221 & 0.0222 & 0.0225 & 0.0298 & 0.0197 & 0.0232 \\
    PSNR  $\textcolor{MyGreen}{\uparrow}$  & 17.0758 & 17.0694 & 17.0521 & 16.9704 & 15.7044 & 17.6116 & 16.8371 \\
    LPIPS $\textcolor{MyRed}{\downarrow}$  & 0.3704 & 0.3796 & 0.3810 & 0.3796 & 0.4011 & 0.3663 & 0.3852 \\
    \bottomrule
  \end{tabular}
  \end{adjustbox}
\end{table}

\textbf{Reconstruction Performance against Defense Mechanisms.}
We evaluate the reconstruction performance of IGF against five common defense mechanisms. The technical details of these common defense mechanisms are introduced in Appendix~\ref{commondefense}, and the reconstruction performance is shown in Table~\ref{defensetable}. When tested against Gradient Pruning with hyperparameters set to $\{0.7, 0.8, 0.9\}$, our method maintains consistent performance with MSE values of $0.0221$, PSNR above $17$, and LPIPS around $0.38$. Against Sign Compression, which quantizes gradients to their signs, our method maintains stable performance, achieving MSE values of approximately $0.0221$, PSNR above $17$, and LPIPS around $0.38$. When confronted with Gaussian Noise, our method still achieves reasonable reconstruction quality (MSE = $0.0225$, PSNR = $15.7044$), though with some performance degradation. This result stems from our learning-based inversion model, which has a strong mapping capability. This demonstrates significant resilience of IGF and allows IGF to almost ignore the defense mechanism and reconstruct the forgotten data. This also highlights the need for a novel defense method that can fundamentally disrupt the attacker's ability to reconstruct meaningful data. 

\begin{figure}[!htbp]
    \centering
    \includegraphics[width=1.0\linewidth]{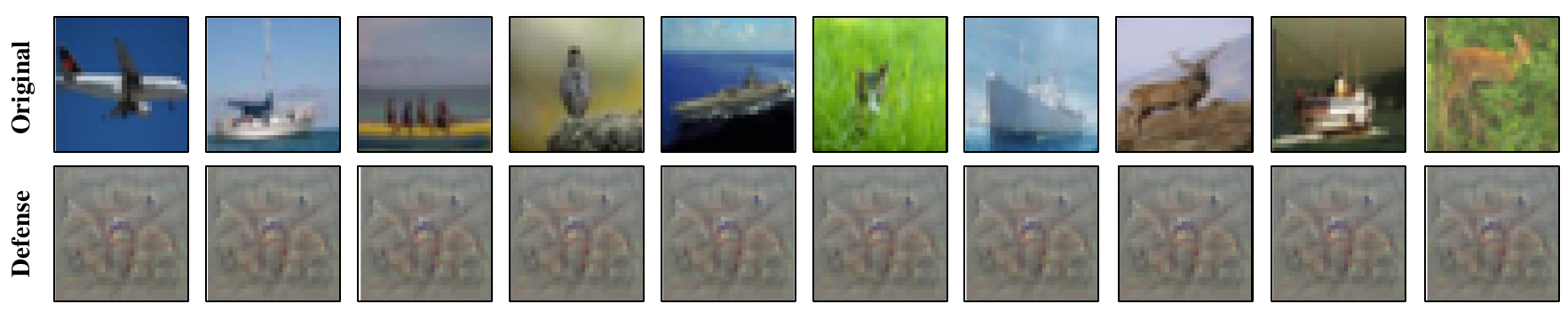}
    \caption{Forgotten images and our reconstructed images on the CIFAR-10 dataset under Orthogonal Obfuscation defense.}
    \label{our_defense}
    \vspace{-1 em}
\end{figure}

\textbf{Reconstruction Performance against Orthogonal Obfuscation Defense.}
As shown in the Figure~\ref{our_defense}, our proposed Orthogonal Obfuscation defense disrupts reconstruction by altering gradient difference directions while preserving their L2-norm. Reconstructed images exhibit random noise, effectively thwarting IGF and protecting sensitive data.  

\subsection{Ablation Studies}

\begin{figure}[!htbp]
    \centering
    \includegraphics[width=\linewidth]{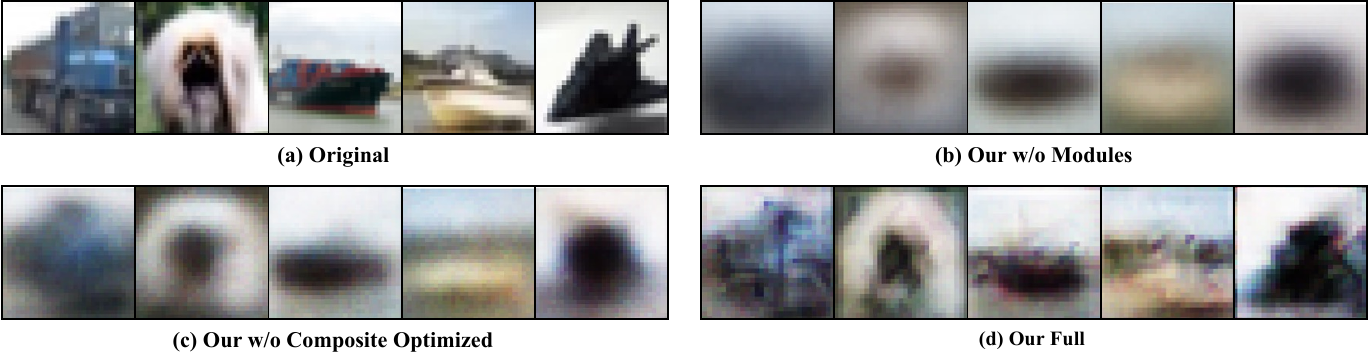}
    \caption{Forgotten images and our reconstructed images using our inversion model across different configurations.}
    \label{fig:ablation}
    \vspace{-1 em}
\end{figure}

To evaluate the contribution of each component in our attack framework, we conduct ablation studies, visualizing reconstruction results under various configurations, as shown in Figure~\ref{fig:ablation}.
\textit{Original} row shows the ground-truth forgotten samples. \textit{Our w/o Modules} configuration, which excludes all proposed modules, exhibits severe degradation in reconstructed images, with prominent artifacts and loss of structural details. This underscores the inherent challenges of reconstruction attacks and the necessity of our enhancements. \textit{Our w/o Composite Optimized} row, which excludes our composite loss-optimization module, produces images that preserve basic shapes but suffer from blurring, color inconsistencies, and a lack of fine details. This highlights the critical role of perceptual loss in capturing high-level semantic features beyond mere pixel-level reconstruction. In contrast, our complete model (\textit{Our Full}), incorporating all proposed components, achieves reconstructions with significantly improved visual quality. These images exhibit sharper definitions, better preservation of textures, and more accurate color reproduction. By effectively balancing low-level pixel information and high-level semantic features, our comprehensive approach yields reconstructions that closely resemble the original forgotten samples. \textbf{Further ablation studies on federated aggregation methods, auxiliary datasets, dimensionality reduction techniques, and the hyperparameter $\beta$ are provided in Appendix~\ref{Additional ablation study}}.


\section{Conclusion}
In this paper, we expose a critical privacy vulnerability in FU by proposing a novel reconstruction attack that exploits gradient differences used as PoFU. Our proposed IGF leverages the latent correlations between gradient differences and forgotten samples to reconstruct large-scale private data from individual PoFU. Through extensive experiments, we demonstrate that our attack achieves high-fidelity reconstruction, exposing the inadequacy of existing FU safeguards. To counter this threat, we introduce an orthogonal obfuscation defense that disrupts the reconstruction process, forcing inverted images into fixed noise patterns that resist recovery. Our findings underscore the fragility of current FU mechanisms against gradient-based and gradient-difference-based attacks, highlighting the urgent need for robust defenses and motivating further exploration of secure FU strategies. 

\newpage
\bibliography{simple}

\begin{thebibliography}{10}

\bibitem{rosen2011right}
Jeffrey Rosen.
\newblock The right to be forgotten.
\newblock {\em Stan. L. Rev. Online}, 64:88, 2011.

\bibitem{pardau2018california}
Stuart~L Pardau.
\newblock The california consumer privacy act: Towards a european-style privacy
  regime in the united states.
\newblock {\em J. Tech. L. \& Pol'y}, 23:68, 2018.

\bibitem{liu2020federated}
Gaoyang Liu, Xiaoqiang Ma, Yang Yang, Chen Wang, and Jiangchuan Liu.
\newblock Federated unlearning.
\newblock {\em arXiv preprint arXiv:2012.13891}, 2020.

\bibitem{lindstrom2024federated}
Cassandra Lindstrom.
\newblock Federated unlearning in financial applications.
\newblock {\em preprints202409.1816}, 2024.

\bibitem{gao2024verifi}
Xiangshan Gao, Xingjun Ma, Jingyi Wang, Youcheng Sun, Bo~Li, Shouling Ji, Peng
  Cheng, and Jiming Chen.
\newblock Verifi: Towards verifiable federated unlearning.
\newblock {\em IEEE Transactions on Dependable and Secure Computing}, 2024.

\bibitem{10416191}
Jiasi Weng, Shenglong Yao, Yuefeng Du, Junjie Huang, Jian Weng, and Cong Wang.
\newblock Proof of unlearning: Definitions and instantiation.
\newblock {\em IEEE Transactions on Information Forensics and Security},
  19:3309--3323, 2024.

\bibitem{10937129}
Xuhan Zuo, Minghao Wang, Tianqing Zhu, Lefeng Zhang, Shui Yu, and Wanlei Zhou.
\newblock Federated learning with blockchain-enhanced machine unlearning: A
  trustworthy approach.
\newblock {\em IEEE Transactions on Services Computing}, pages 1--15, 2025.

\bibitem{salem2020updates}
Ahmed Salem, Apratim Bhattacharya, Michael Backes, Mario Fritz, and Yang Zhang.
\newblock $\{$Updates-Leak$\}$: Data set inference and reconstruction attacks
  in online learning.
\newblock In {\em 29th USENIX security symposium (USENIX Security 20)}, pages
  1291--1308, 2020.

\bibitem{wu2022federated}
Chen Wu, Sencun Zhu, and Prasenjit Mitra.
\newblock Federated unlearning with knowledge distillation.
\newblock {\em arXiv preprint arXiv:2201.09441}, 2022.

\bibitem{wang2022federated}
Junxiao Wang, Song Guo, Xin Xie, and Heng Qi.
\newblock Federated unlearning via class-discriminative pruning.
\newblock In {\em Proceedings of the ACM web conference 2022}, pages 622--632,
  2022.

\bibitem{zhong2025unlearning}
Zhengyi Zhong, Weidong Bao, Ji~Wang, Shuai Zhang, Jingxuan Zhou, Lingjuan Lyu,
  and Wei Yang~Bryan Lim.
\newblock Unlearning through knowledge overwriting: Reversible federated
  unlearning via selective sparse adapter.
\newblock {\em arXiv preprint arXiv:2502.20709}, 2025.

\bibitem{boenisch2023curious}
Franziska Boenisch, Adam Dziedzic, Roei Schuster, Ali~Shahin Shamsabadi, Ilia
  Shumailov, and Nicolas Papernot.
\newblock When the curious abandon honesty: Federated learning is not private.
\newblock In {\em 2023 IEEE 8th European Symposium on Security and Privacy
  (EuroS\&P)}, pages 175--199. IEEE, 2023.

\bibitem{le2023privacy}
Junqing Le, Di~Zhang, Xinyu Lei, Long Jiao, Kai Zeng, and Xiaofeng Liao.
\newblock Privacy-preserving federated learning with malicious clients and
  honest-but-curious servers.
\newblock {\em IEEE Transactions on Information Forensics and Security},
  18:4329--4344, 2023.

\bibitem{zhu2019deep}
Ligeng Zhu, Zhijian Liu, and Song Han.
\newblock Deep leakage from gradients.
\newblock {\em Advances in neural information processing systems}, 32, 2019.

\bibitem{geiping2020inverting}
Jonas Geiping, Hartmut Bauermeister, Hannah Dr{\"o}ge, and Michael Moeller.
\newblock Inverting gradients-how easy is it to break privacy in federated
  learning?
\newblock {\em Advances in neural information processing systems},
  33:16937--16947, 2020.

\bibitem{hu2024learn}
Hongsheng Hu, Shuo Wang, Tian Dong, and Minhui Xue.
\newblock Learn what you want to unlearn: Unlearning inversion attacks against
  machine unlearning.
\newblock In {\em 2024 IEEE Symposium on Security and Privacy (SP)}, pages
  3257--3275. IEEE, 2024.

\bibitem{279996}
Anvith Thudi, Hengrui Jia, Ilia Shumailov, and Nicolas Papernot.
\newblock On the necessity of auditable algorithmic definitions for machine
  unlearning.
\newblock In {\em 31st USENIX Security Symposium (USENIX Security 22)}, pages
  4007--4022, Boston, MA, August 2022. USENIX Association.

\bibitem{li2022auditing}
Zhuohang Li, Jiaxin Zhang, Luyang Liu, and Jian Liu.
\newblock Auditing privacy defenses in federated learning via generative
  gradient leakage.
\newblock In {\em Proceedings of the IEEE/CVF Conference on Computer Vision and
  Pattern Recognition}, pages 10132--10142, 2022.

\bibitem{liu2022right}
Yi~Liu, Lei Xu, Xingliang Yuan, Cong Wang, and Bo~Li.
\newblock The right to be forgotten in federated learning: An efficient
  realization with rapid retraining.
\newblock In {\em IEEE INFOCOM 2022-IEEE conference on computer
  communications}, pages 1749--1758. IEEE, 2022.

\bibitem{halimi2207federated}
Anisa Halimi, Swanand Kadhe, Ambrish Rawat, and Nathalie Baracaldo.
\newblock Federated unlearning: How to efficiently erase a client in fl?, 2022.
\newblock {\em URL https://arxiv. org/abs/2207.05521}, 2022.

\bibitem{wang2024fedu}
Weiqi Wang, Chenhan Zhang, Zhiyi Tian, and Shui Yu.
\newblock Fedu: Federated unlearning via user-side influence approximation
  forgetting.
\newblock {\em IEEE Transactions on Dependable and Secure Computing}, 2024.

\bibitem{xu2024update}
Heng Xu, Tianqing Zhu, Lefeng Zhang, Wanlei Zhou, and S~Yu Philip.
\newblock Update selective parameters: Federated machine unlearning based on
  model explanation.
\newblock {\em IEEE Transactions on Big Data}, 2024.

\bibitem{gu2024unlearning}
Hanlin Gu, Gongxi Zhu, Jie Zhang, Xinyuan Zhao, Yuxing Han, Lixin Fan, and
  Qiang Yang.
\newblock Unlearning during learning: An efficient federated machine unlearning
  method.
\newblock {\em arXiv preprint arXiv:2405.15474}, 2024.

\bibitem{chen2025fedmua}
Jian Chen, Zehui Lin, Wanyu Lin, Wenlong Shi, Xiaoyan Yin, and Di~Wang.
\newblock Fedmua: Exploring the vulnerabilities of federated learning to
  malicious unlearning attacks.
\newblock {\em IEEE Transactions on Information Forensics and Security}, 2025.

\bibitem{wang2025poisoning}
Wenbin Wang, Qiwen Ma, Zifan Zhang, Yuchen Liu, Zhuqing Liu, and Minghong Fang.
\newblock Poisoning attacks and defenses to federated unlearning.
\newblock {\em arXiv preprint arXiv:2501.17396}, 2025.

\bibitem{zhang2023generative}
Chi Zhang, Zhang Xiaoman, Ekanut Sotthiwat, Yanyu Xu, Ping Liu, Liangli Zhen,
  and Yong Liu.
\newblock Generative gradient inversion via over-parameterized networks in
  federated learning.
\newblock In {\em Proceedings of the IEEE/CVF International Conference on
  Computer Vision}, pages 5126--5135, 2023.

\bibitem{jeon2021gradient}
Jinwoo Jeon, Kangwook Lee, Sewoong Oh, Jungseul Ok, et~al.
\newblock Gradient inversion with generative image prior.
\newblock {\em Advances in neural information processing systems},
  34:29898--29908, 2021.

\bibitem{fang2023gifd}
Hao Fang, Bin Chen, Xuan Wang, Zhi Wang, and Shu-Tao Xia.
\newblock Gifd: A generative gradient inversion method with feature domain
  optimization.
\newblock In {\em Proceedings of the IEEE/CVF International Conference on
  Computer Vision}, pages 4967--4976, 2023.

\bibitem{sun2024gi}
Yu~Sun, Gaojian Xiong, Xianxun Yao, Kailang Ma, and Jian Cui.
\newblock Gi-pip: Do we require impractical auxiliary dataset for gradient
  inversion attacks?
\newblock In {\em ICASSP 2024-2024 IEEE International Conference on Acoustics,
  Speech and Signal Processing (ICASSP)}, pages 4675--4679. IEEE, 2024.

\bibitem{wu2023learning}
Ruihan Wu, Xiangyu Chen, Chuan Guo, and Kilian~Q Weinberger.
\newblock Learning to invert: Simple adaptive attacks for gradient inversion in
  federated learning.
\newblock In {\em Uncertainty in Artificial Intelligence}, pages 2293--2303.
  PMLR, 2023.

\bibitem{mcmahan2017communication}
Brendan McMahan, Eider Moore, Daniel Ramage, Seth Hampson, and Blaise~Aguera
  y~Arcas.
\newblock Communication-efficient learning of deep networks from decentralized
  data.
\newblock In {\em Artificial intelligence and statistics}, pages 1273--1282.
  PMLR, 2017.

\bibitem{orthogonalization_eom}
Orthogonalization.
\newblock In {\em Encyclopedia of Mathematics}. EMS Press, 2001.
\newblock \url{https://encyclopediaofmath.org/wiki/Orthogonalization}.

\bibitem{krizhevsky2009learning}
Alex Krizhevsky, Geoffrey Hinton, et~al.
\newblock Learning multiple layers of features from tiny images.
\newblock {\em Placeholder Journal}, 2009.

\bibitem{lecun1998gradient}
Yann LeCun, L{\'e}on Bottou, Yoshua Bengio, and Patrick Haffner.
\newblock Gradient-based learning applied to document recognition.
\newblock {\em Proceedings of the IEEE}, 86(11):2278--2324, 1998.

\bibitem{xiao2017fashion}
Han Xiao, Kashif Rasul, and Roland Vollgraf.
\newblock Fashion-mnist: a novel image dataset for benchmarking machine
  learning algorithms.
\newblock {\em arXiv preprint arXiv:1708.07747}, 2017.

\bibitem{zhang2018unreasonable}
Richard Zhang, Phillip Isola, Alexei~A Efros, Eli Shechtman, and Oliver Wang.
\newblock The unreasonable effectiveness of deep features as a perceptual
  metric.
\newblock In {\em Proceedings of the IEEE conference on computer vision and
  pattern recognition}, pages 586--595, 2018.

\bibitem{he2016deep}
Kaiming He, Xiangyu Zhang, Shaoqing Ren, and Jian Sun.
\newblock Deep residual learning for image recognition.
\newblock In {\em Proceedings of the IEEE conference on computer vision and
  pattern recognition}, pages 770--778, 2016.

\bibitem{li2020federated}
Tian Li, Anit~Kumar Sahu, Manzil Zaheer, Maziar Sanjabi, Ameet Talwalkar, and
  Virginia Smith.
\newblock Federated optimization in heterogeneous networks.
\newblock {\em Proceedings of Machine learning and systems}, 2:429--450, 2020.

\bibitem{reddi2020adaptive}
Sashank Reddi, Zachary Charles, Manzil Zaheer, Zachary Garrett, Keith Rush,
  Jakub Kone{\v{c}}n{\`y}, Sanjiv Kumar, and H~Brendan McMahan.
\newblock Adaptive federated optimization.
\newblock {\em arXiv preprint arXiv:2003.00295}, 2020.

\bibitem{weinberger2009feature}
Kilian Weinberger, Anirban Dasgupta, John Langford, Alex Smola, and Josh
  Attenberg.
\newblock Feature hashing for large scale multitask learning.
\newblock In {\em Proceedings of the 26th annual international conference on
  machine learning}, pages 1113--1120, 2009.

\end{thebibliography}
\bibliographystyle{unsrt}

\newpage
\appendix

\section{Discussion and Limitations}

To the best of our knowledge, IGF framework is the first to exploit gradient differences as an attack surface in federated unlearning (FU). Previous reconstruction attacks in machine learning and federated learning (FL)~\cite{zhu2019deep, geiping2020inverting, sun2024gi, wu2023learning} directly leverage sample-level gradients, which inherently contain richer sample information. Currently, the only available baseline for reconstruction attacks in FU is GIAMU~\cite{hu2024learn}, which relies on white-box access to both the original and unlearned models and overlooks the privacy vulnerabilities arising from gradient differences sharing during verification. Our work addresses this gap by demonstrating that an \textit{honest-but-curious} adversary with partial prior knowledge can reconstruct forgotten samples by inverting gradient differences. Additionally, in extreme scenarios, such as a black-box setting where the adversary lacks prior knowledge or cannot exploit the directionality of gradient differences, the attack's complexity increases significantly, making the reconstruction of forgotten data largely unexplored.


\section{Experimental Settings}
\begin{table}[ht]
  \caption{Mathematical notations}
  \label{table:notations}
  \centering
  \begin{tabular}{cc}
    \toprule
    \textbf{Notation} & \textbf{Description}                                                                           \\ \hline
    $\mathcal{C}$        & \multicolumn{1}{c}{Set of Clients} \\
    $\mathcal{D}$        & \multicolumn{1}{c}{Global Dataset}      \\
    $H$         & \multicolumn{1}{c}{The Number of Clients}                                                                    \\
    $\mathbf{M}$        & \multicolumn{1}{c}{Original Global Model}                                                                             \\
    $^u\mathbf{M}$        & \multicolumn{1}{c}{Unlearned Global Model}                                                                             \\
    $\mathbf{g}$        & \multicolumn{1}{c}{Stochastic Gradient} \\
    $\mathcal{G}$        & \multicolumn{1}{c}{Gradient Queried by Adversary} \\
    $\ell$        & \multicolumn{1}{c}{Loss Function in Local Training} \\
    $T$        & \multicolumn{1}{c}{Number of Global Rounds} \\
    $\mathcal{I}$        & \multicolumn{1}{c}{Inversion Model} \\
    $(x,y)$        & \multicolumn{1}{c}{Data Point} \\
    $B$        & \multicolumn{1}{c}{Batch Size} \\
    $\phi(\cdot)$        & \multicolumn{1}{c}{Intermediate Feature Extractor} \\
    $\mathbf{\Delta} \mathcal{G}, \mathbf{\Delta} \mathbf{g}$        & \multicolumn{1}{c}{Gradient Differences} \\
    $d$        & \multicolumn{1}{c}{Model Size} \\
    $\textbf{U}$        & \multicolumn{1}{c}{Left Singular Vectors} \\
    $\textbf{V}$        & \multicolumn{1}{c}{Right Singular Vectors} \\
    $\mathbf{r}$        & \multicolumn{1}{c}{Random Vector} \\
    $\mathcal{M}$        & \multicolumn{1}{c}{Mask Matrix in Gradient Pruning} \\
    $\epsilon$        & \multicolumn{1}{c}{Gaussian Noise} \\
    $w$        & \multicolumn{1}{c}{Window Size in Gradient Smoothing} \\
    $\zeta$        & \multicolumn{1}{c}{Parameter Deviation Constraint Radius} \\
    $\boldsymbol{\Psi}$        & \multicolumn{1}{c}{Gradient Differences Matrix} \\
    $\mathbf{V}^{[k]}$ & \multicolumn{1}{c}{Projection Matrix} \\
    \bottomrule
  \end{tabular}
\end{table}

\subsection{Datasets}
\label{Appendix_Datasets}
We evaluate IGF using the widely adopted CIFAR-10 and CIFAR-100 datasets~\cite{krizhevsky2009learning}, both standard benchmarks in reconstruction attacks and federated learning research. CIFAR-10 comprises 60,000 color images ($32\times32$ pixels) across 10 categories, with 50,000 images for training and 10,000 for testing. Each category contains 6,000 images. CIFAR-100 is structured similarly but includes 100 categories, each with 600 images, totaling 60,000 images (50,000 for training and 10,000 for testing).

Additionally, we assess our approach on the MNIST~\cite{lecun1998gradient} and Fashion-MNIST datasets~\cite{xiao2017fashion}. MNIST consists of 70,000 grayscale images ($28\times28$ pixels) of handwritten digits, divided into 60,000 training and 10,000 testing images. Fashion-MNIST, designed as a more challenging alternative, also contains 70,000 $28\times28$ grayscale images but represents 10 categories of fashion items. It mirrors MNIST's training and testing split.

\subsection{Details of Metrics}
\label{appendix_metrics}

MSE measures the average squared difference between the original forgotten image and the reconstructed image. It is widely used as a loss function in image processing tasks and image quality assessment $\text{MSE} = \frac{1}{N} \sum_{i=1}^{N} (x_i - \hat{x}_i)^2$, where $x_i$ is the pixel value of the original forgotten image and $\hat{x}_i$ is the pixel value of the reconstruction.

PSNR measures the quality of the reconstructed or compressed image relative to the forgotten image. It is expressed in decibels (dB) and is inversely related to MSE—lower MSE values correspond to higher PSNR values. $\text{PSNR} = 10 \cdot \log_{10} \left( \frac{R^2}{\text{MSE}} \right)$, Where \( R \) is the maximum pixel value.

LPIPS~\cite{zhang2018unreasonable} is a perceptual similarity metric designed to assess the perceptual quality of images based on learned features from a neural network (typically a pretrained deep network like VGG). Unlike MSE and PSNR, LPIPS is more aligned with human visual perception, focusing on perceptual similarity rather than pixel-level accuracy
$\text{LPIPS}(x, \hat{x}) = \frac{1}{L} \sum_{l=1}^{L} \left\| \phi_l(x) - \phi_l(\hat{x}) \right\|_2^2$, where \( L \) is the total number of layers used for feature extraction. \( \| \cdot \|_2 \) is the Euclidean distance (L2 norm) between the feature maps.

\subsection{Details of Common Defense Mechanisms}\label{commondefense}

This section outlines five defense mechanisms~\cite{wu2023learning} designed to obfuscate shared gradients and mitigate gradient-based reconstruction attacks through various perturbation techniques. Given an input gradient vector $\mathbf{g}$, each mechanism produces an obfuscated gradient vector $\mathbf{g}'$. We adapt these mechanisms to perturb shared gradient differences in FU.

(a) \textbf{Sign Compression.} The sign compression mechanism applies the $\mathsf{sign}$ operation to each component of the gradient $\mathbf{g}$, retaining only its sign ($-1$, $0$, or $1$) and discarding magnitude information. This preserves the gradient's direction while significantly reducing communication overhead, as only sign bits are transmitted. By limiting the attacker's access to sign information, this method increases the difficulty of reconstructing forgotten data. The operation is defined as:
\begin{equation}
\mathbf{g}' = \mathsf{sign}(\mathbf{g}), \quad \text{where} \quad \mathsf{sign}(\mathbf{g}_i) =
\begin{cases}
1, & \text{if } \mathbf{g}_i > 0 \\
-1, & \text{if } \mathbf{g}_i < 0 \\
0, & \text{if } \mathbf{g}_i = 0
\end{cases}
\end{equation}

(b) \textbf{Gradient Pruning.} Gradient pruning sparsifies the gradient by retaining only the $k$ components with the largest absolute values, setting all others to zero. A binary mask $\mathcal{M}$ selectively preserves these significant components. Widely used in FL to reduce communication costs, this method also enhances privacy by limiting the attacker's access to a subset of gradient components, complicating the inference of forgotten data. The operation is formulated as:
\begin{equation}
\mathbf{g}' = \mathbf{g} \odot \mathcal{M},
\end{equation}
where $\odot$ denotes element-wise multiplication, and $\mathcal{M}$ is the mask matrix.

(c) \textbf{Gaussian Noise.} This mechanism perturbs the gradient $\mathbf{g}$ by adding independent and identically distributed Gaussian noise $\epsilon \sim \mathcal{N}(0, \sigma^2 \mathbf{I})$. Controlled by the standard deviation $\sigma$, the noise introduces uncertainty to achieve differential privacy, obscuring precise gradient values and hindering reconstruction of forgotten data. The operation is expressed as:
\begin{equation}
\mathbf{g}' = \mathbf{g} + \epsilon, \quad \epsilon \sim \mathcal{N}(0, \sigma^2 \mathbf{I}).
\end{equation}

(d) \textbf{Gradient Perturbation.} This method perturbs the gradient by adding noise proportional to the gradient's magnitude, applying larger perturbations to dimensions with greater gradient values. The perturbed gradient is defined as:
\begin{equation}
\mathbf{g}' = \mathbf{g} + \left( \mathcal{N}(\mathbf{0}, \mathbf{I}) \times \text{scale} \right) \times \left( |\mathbf{g}| \times \text{factor} \right),
\end{equation}
where $\mathcal{N}(\mathbf{0}, \mathbf{I})$ is a standard normal random tensor, $\text{scale}$ determines the base perturbation magnitude, and $\text{factor}$ adjusts the sensitivity of the perturbation to the gradient's amplitude.

(e) \textbf{Gradient Smoothing.} Gradient smoothing mitigates high-frequency variations in the gradient by applying a moving average over the feature dimensions, blending the result with the original gradient. The operation is formulated as:
\begin{equation}
\mathbf{g}' = \text{reshape} \left( (1 - \alpha_{\text{gs}}) \mathbf{g}^{\text{flat}} + \alpha_{\text{gs}} \cdot \text{MA}_w(\mathbf{g}^{\text{flat}}) \right),
\end{equation}
where $\mathbf{g}^{\text{flat}}$ is the flattened gradient, $\text{MA}_w$ denotes the moving average with window size $w$, and $\alpha_{\text{gs}} \in [0, 1]$ controls the smoothing intensity.

\section{Additional Ablation Studies}
\label{Additional ablation study}

\subsection{Impact of Different Federated Aggregation Methods}

\begin{figure}[!htbp]
    \centering
    \includegraphics[width=0.8\linewidth]{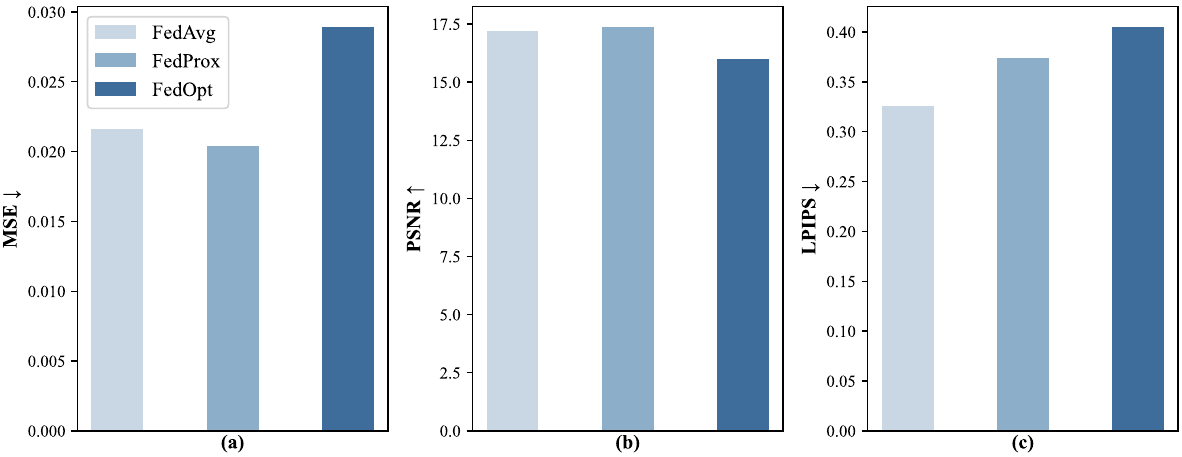}
    \caption{The reconstruction performance under different federated aggregation methods.}
    \label{federated-agg}
\end{figure}

We investigated how three federated aggregation methods, including FedAvg~\cite{mcmahan2017communication}, FedProx~\cite{li2020federated}, and FedOpt~\cite{reddi2020adaptive}, affect the p of reconstruction attacks in FU scenarios. Figure~\ref{federated-agg} illustrates the performance of our attack method across various aggregation algorithms commonly used in FL systems. The results demonstrate that while aggregation methods can influence reconstruction quality, our attack remains effective across different techniques. When examining more sophisticated aggregation methods like FedProx and FedOpt, we observe slightly different reconstruction patterns, but the overall attack effectiveness remains consistent.

\subsection{Impact of Different Distributions of Auxiliary Datasets}

\begin{table}[ht]
\centering
\renewcommand\arraystretch{1.5}
\caption{The reconstruction performance of different distributions of auxiliary datasets.}
\begin{tabular}{cccc}
\toprule
\textbf{Distribution of Auxiliary Datasets}  & \textbf{MSE} $\textcolor{MyRed}{\downarrow}$     & \textbf{PSNR} $\textcolor{MyGreen}{\uparrow}$   & \textbf{LPIPS} $\textcolor{MyRed}{\downarrow}$   \\ \hline
In-Distribution     & 0.0211 & 17.1947 & 0.3261 \\
Out-of-Distribution & 0.0259 & 16.6364 & 0.3324 \\ \bottomrule
\end{tabular}
\label{outofdis}
\end{table}

In real-world scenarios, adversaries often struggle to obtain the complete data distribution of clients. To investigate the effectiveness of IGF attacks under entirely different distributions, we partition the CIFAR-10 training dataset as the federated client dataset and employ CIFAR-100, comprising entirely different categories, as auxiliary data. This setup simulates a reconstruction attack where the adversary lacks knowledge of the client data distribution. As demonstrated in Table~\ref{outofdis}, the IGF attack retains strong efficacy even with out-of-distribution auxiliary data, exhibiting only marginal degradation across various performance metrics.

\begin{figure}[!htbp]
    \centering
    \includegraphics[width=0.8\linewidth]{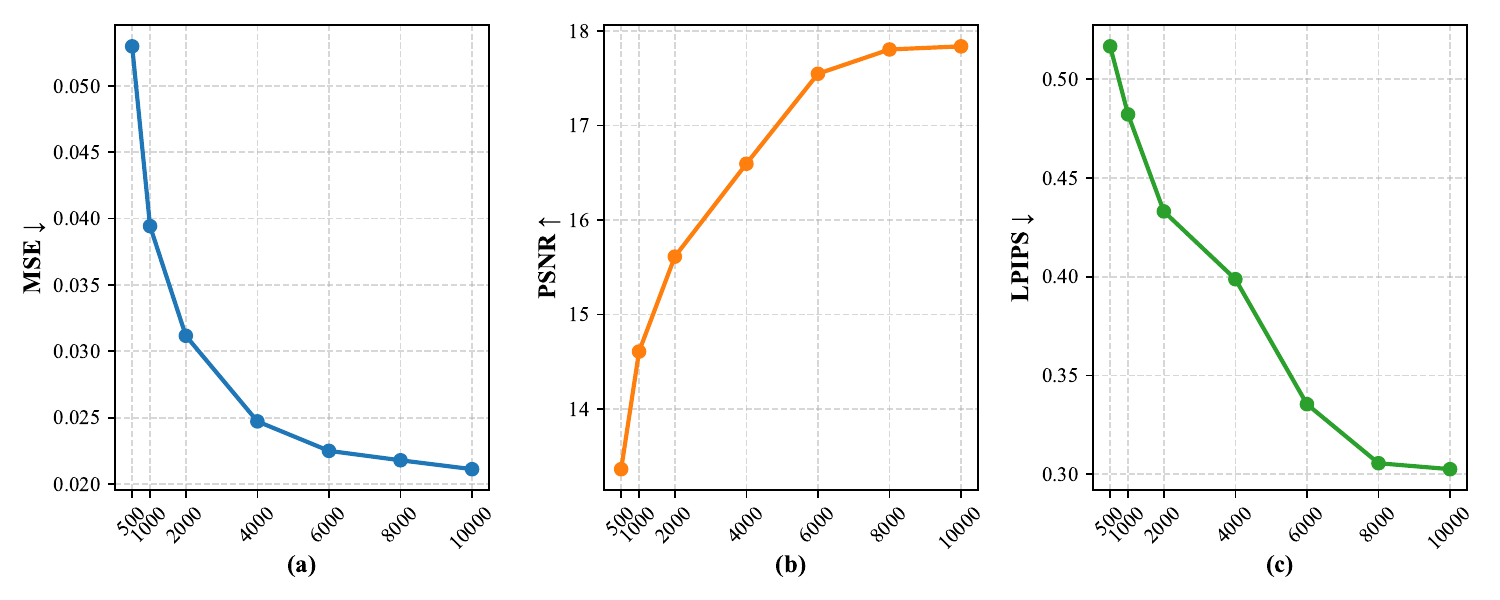}
    \caption{The reconstruction performance with different auxiliary dataset sizes.}
    \label{auxiliary dataset sizes}
\end{figure}

\subsection{Impact of Different Auxiliary Dataset Sizes}
We investigate the influence of varying auxiliary dataset sizes on the efficacy of our attack method. As illustrated in Figure~\ref{auxiliary dataset sizes}, we incrementally scale the dataset from 500 to 10,000 samples. Experimental results reveal that performance metrics stabilize when the auxiliary dataset comprises approximately 6,000 to 8,000 samples, demonstrating that our method achieves efficient and robust performance without requiring extensive auxiliary data. Notably, even with a modest dataset size, our proposed attack method effectively leverages available knowledge to deliver high-quality image reconstruction.

\subsection{Comparative Ablation Study of Dimensionality Reduction Methods}

\begin{figure}[!htbp]
    \centering
    \includegraphics[width=0.7\linewidth]{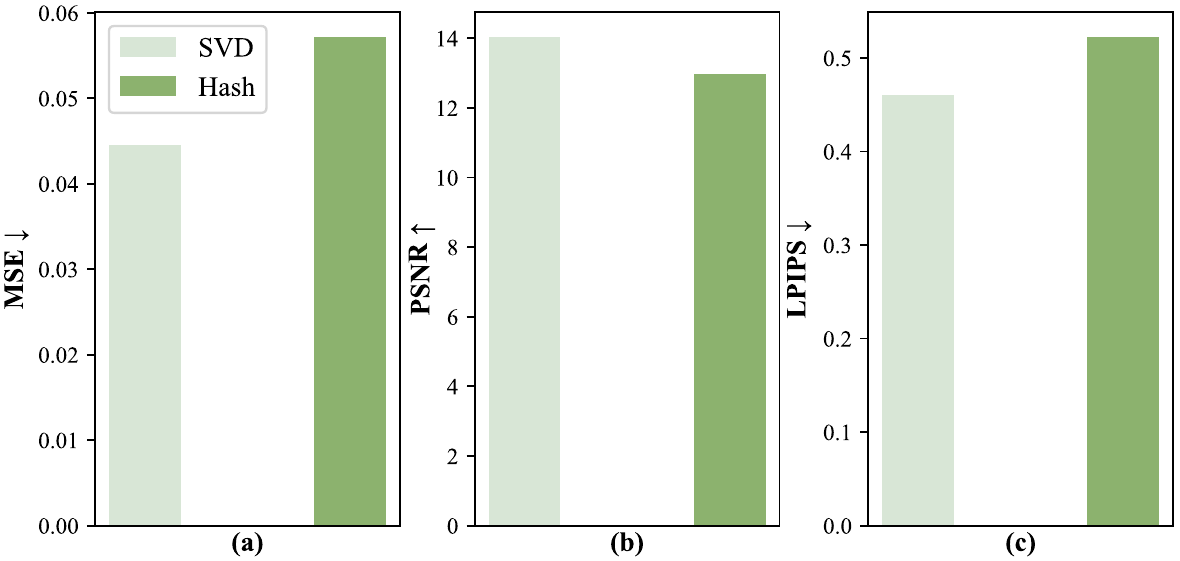}
    \caption{Comparison of the reconstruction effectiveness with applying SVD and Hash dimensionality reduction.}
    \label{SVD_exp}
\end{figure}

To gain a deeper understanding of the effectiveness of dimensionality reduction methods, we compared the performance of Hash-based dimensionality reduction and Singular Value Decomposition (SVD) in terms of reduction quality and reconstruction results. Hash-based dimensionality reduction~\cite{weinberger2009feature} is a vector compression method that relies on random projection, mapping high-dimensional gradient differences to a lower-dimensional space through a sparse random matrix. Specifically, a sparse matrix is constructed where each high-dimensional vector component is randomly assigned to a lower-dimensional target dimension, and each reduced dimension represents the cumulative sum of the corresponding high-dimensional gradient differences. This approach is computationally efficient and well-suited for rapidly compressing gradient differences. However, its randomness disregards the inherent structure of the gradient differences, potentially leading to significant information loss.

\begin{wraptable}{r}{0.32\textwidth}
\centering
\vspace{-3mm}
\begin{tabular}{cc}
\toprule
\textbf{Method}   & \textbf{Size}   \\ \hline
Original & 269722 \\
Hash     & 134861 \\
SVD      & 433    \\ \bottomrule
\end{tabular}
\caption{Comparison of the effectiveness of SVD and Hash for gradient differences reduction.}
\label{tablesvd}
\end{wraptable}
As shown in Figure~\ref{SVD_exp}, SVD outperforms the reconstruction after Hash dimensionality reduction in both reconstruction effects, and as shown in Table~\ref{tablesvd} achieves more significant dimensionality reduction by extracting only key information. SVD-based dimensionality reduction is a data-driven method that decomposes the covariance matrix of the gradient differences to extract principal component directions as the projection basis. SVD dynamically selects the number of dimensions to retain a substantial portion of the variance (e.g., 95\%), ensuring that the reduced results capture the primary patterns of the original gradient differences. 

SVD outperforms Hash-based reduction because it prioritizes the retention of critical information while minimizing the impact of irrelevant noise. Furthermore, in reconstruction tasks, SVD-preserved gradient differences maintain structured features, enabling inversion models to more effectively learn the mapping from lower-dimensional features to the original data, resulting in higher-quality reconstructed images. Conversely, Hash-based reduction disrupts the gradient differences structure through random mixing, making it challenging for reconstruction networks to disentangle useful information, which often leads to blurry or distorted reconstructed images.

\subsection{Comparative Ablation Study of different \texorpdfstring{$\beta$}{beta} parameters}
\begin{figure}
    \centering
    \includegraphics[width=0.8\linewidth]{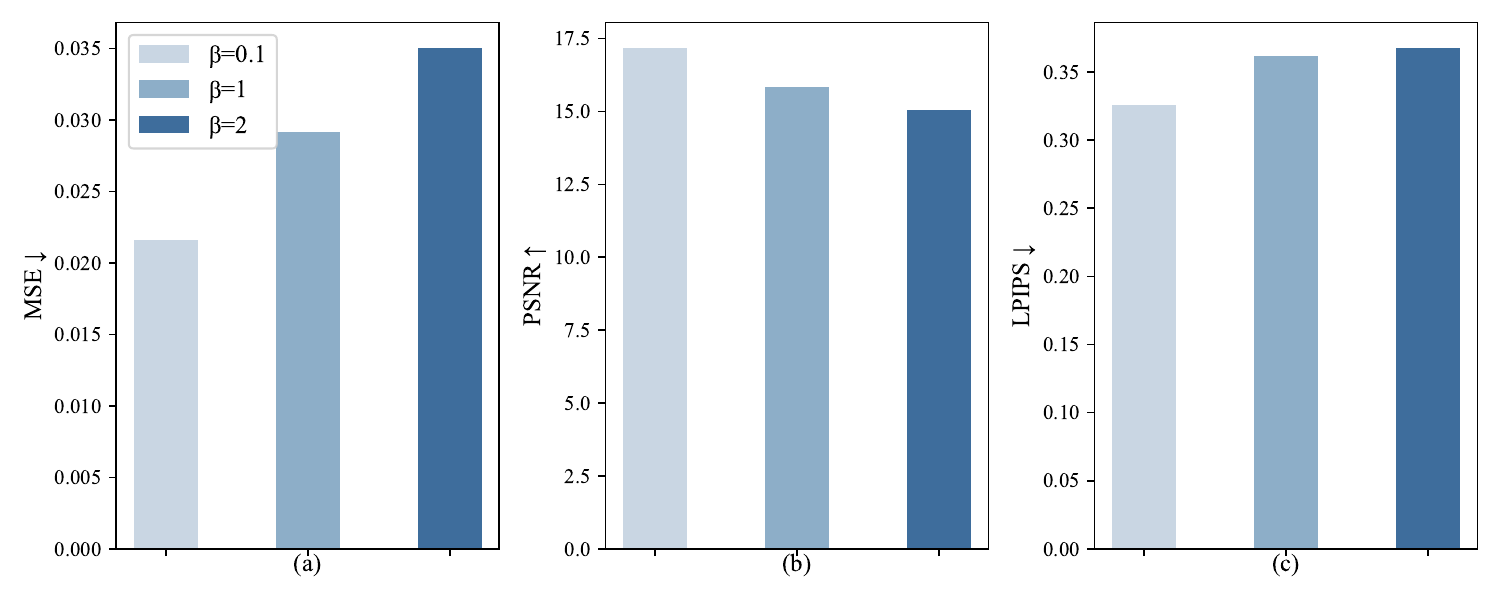}
    \caption{The reconstruction performance under different $\beta$.}
    \label{beta}
\end{figure}

To investigate the role of the parameter $\beta$ in the loss function, which governs the trade-off between pixel-level accuracy and perceptual quality, we conduct an ablation study to assess its impact on reconstruction attack performance. Specifically, we evaluate the effect of varying $\beta \in \{0.1, 1.0, 2.0\}$ on three key metrics: MSE, PSNR, and LPIPS. As shown in Figure~\ref{beta}, increasing $\beta$ reveals a clear trade-off: pixel-level accuracy degrades, as indicated by worsening MSE, and perceptual quality diminishes, as reflected by deteriorating LPIPS, while PSNR exhibits a peak at an intermediate $\beta$ before declining. These findings underscore $\beta$'s critical role in mediating the balance between pixel-wise fidelity and high-level perceptual features.

\section{Inversion Model Architecture}
\begin{figure}[!htbp]
    \centering
    \includegraphics[width=0.7\linewidth]{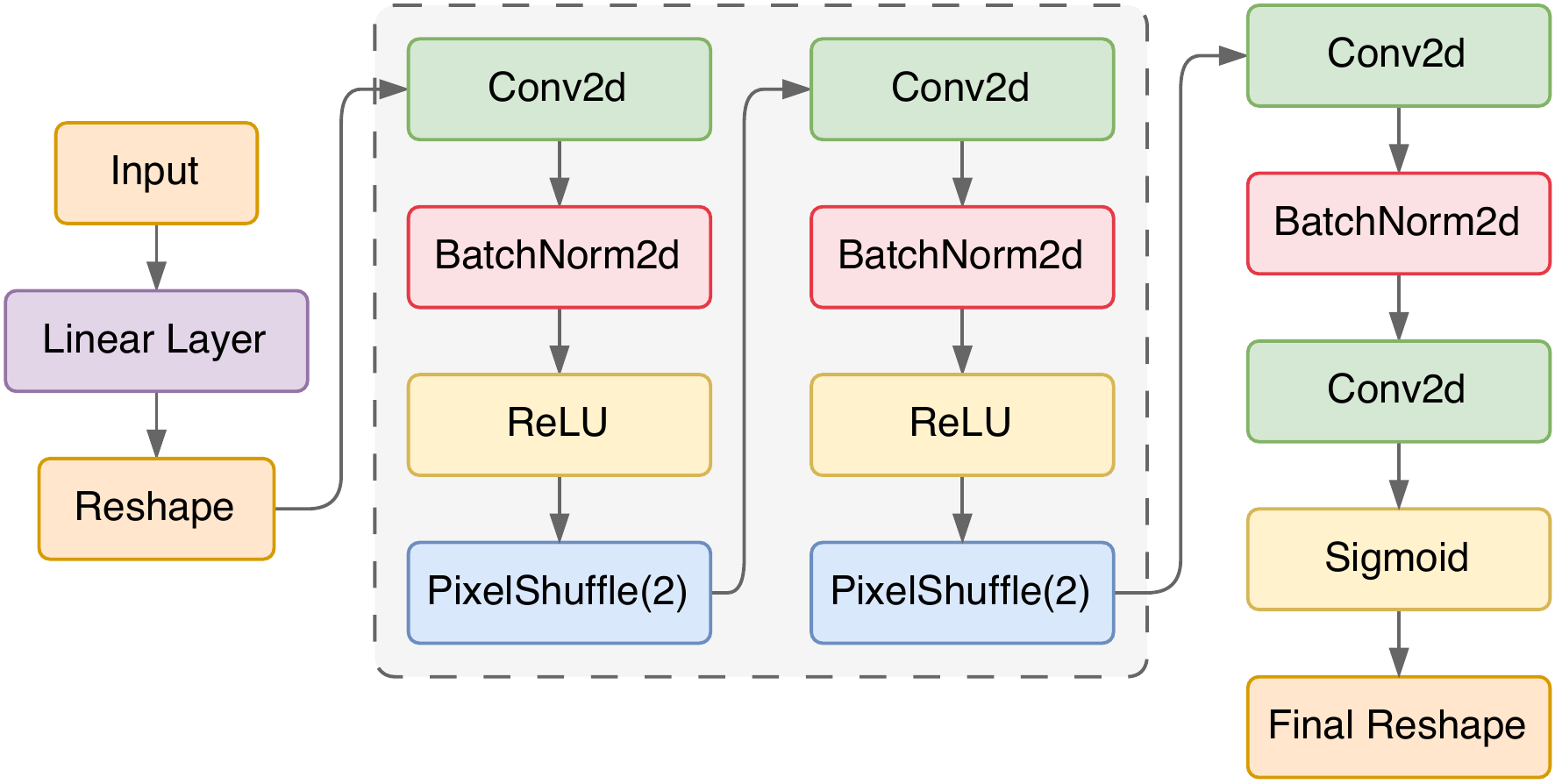}
    \caption{Architecture of the proposed pixel-level inversion model.} 
    \label{inversion_model_fig}
\end{figure}

As illustrated in Figure~\ref{inversion_model_fig}, our pixel-level inversion model features a carefully designed architecture comprising multiple Conv2d and BatchNorm2d layers. We incorporate PixelShuffle for effective upsampling, minimizing artifacts in reconstructed results. A linear layer paired with an initial Reshape operation enhances input processing, while a final Sigmoid activation and Reshape ensure high-quality output generation.

\label{Inversion model architecture}

\section{Additional Reconstructed Images}
\begin{figure}[!htbp]
    \centering
    \includegraphics[width=\linewidth]{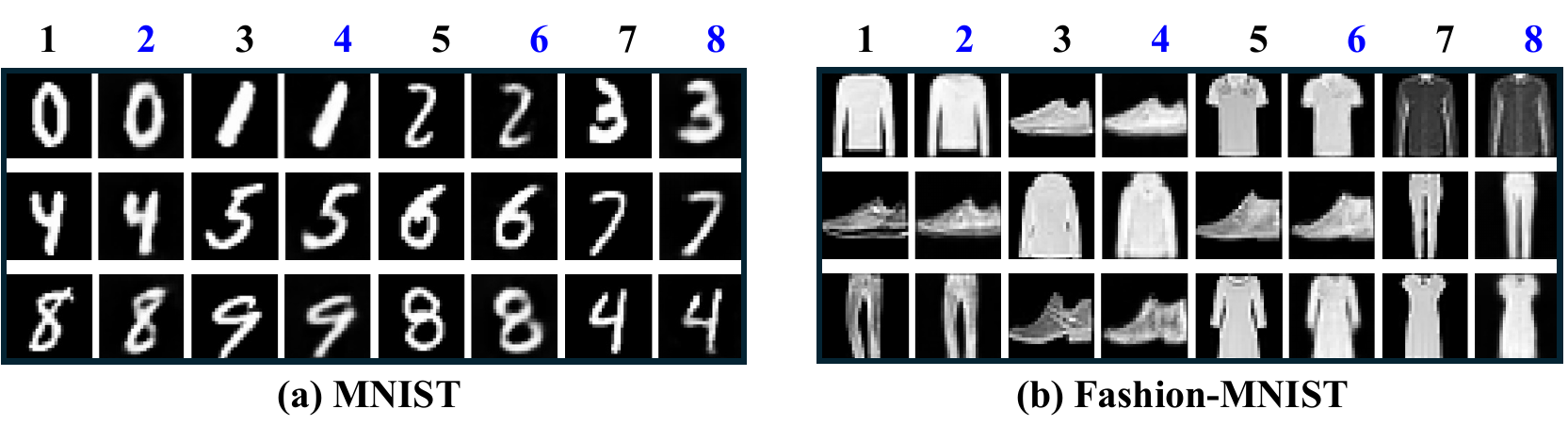}
    \caption{Forgotten and reconstructed images on MNIST and Fashion-MNIST within 1,000 randomly forgotten samples.} 
    \label{reconstruct-mnist}
\end{figure}

\begin{figure}[!htbp]
    \centering
    \includegraphics[width=0.82 \linewidth]{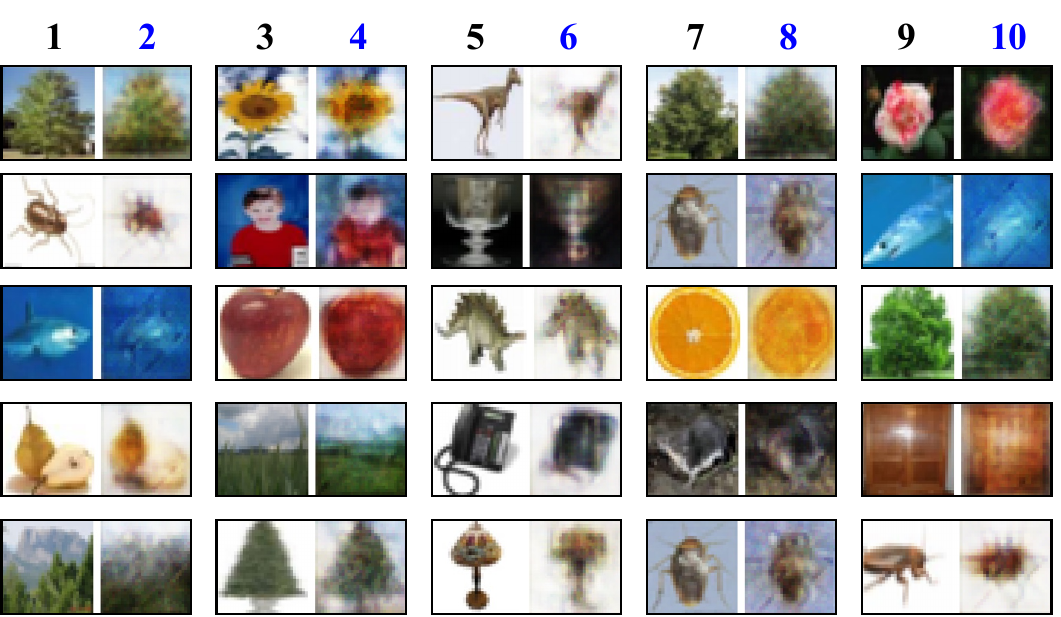}
    \caption{Forgotten and reconstructed images on CIFAR-100.}
    \label{reconstruct-CIFAR100}
\end{figure}

\begin{figure}[!htbp]
    \centering
    \includegraphics[width=0.65\linewidth]{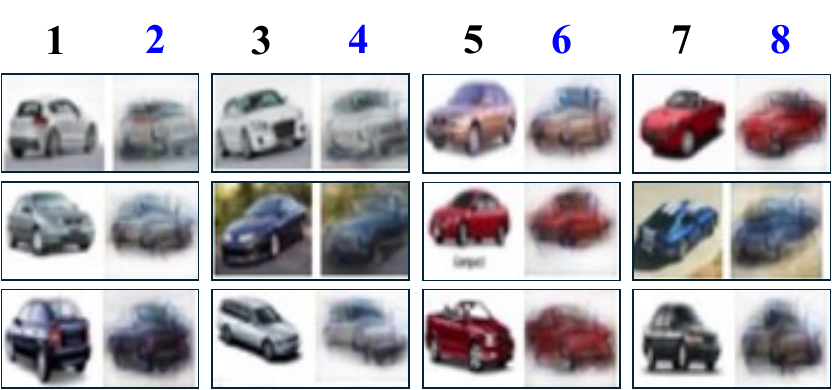}
    \caption{Forgotten and reconstructed images on CIFAR-10 for the unlearned class (car).}
    \label{reconstruct-CIFAR10-car}
\end{figure}

This section showcases the forgotten images and their corresponding reconstructions across multiple datasets, as presented in Figures~\ref{reconstruct-mnist},~\ref{reconstruct-CIFAR100}, and~\ref{reconstruct-CIFAR10-car}. In each figure, \textbf{odd columns display the original images}, and \textcolor{blue!80!white}{\textbf{even columns show our reconstructed results}}. 

For the scenario of \textit{class-level unlearning}, Figure~\ref{reconstruct-CIFAR10-car} presents the forgotten images and reconstruction results on CIFAR-10 for the unlearned class (car). 

\clearpage

\end{document}